\documentclass[pra,aps,twocolumn,floatfix,showpacs]{revtex4}
\usepackage{graphicx,graphics,psfrag,amsmath,calc}
\usepackage{epsfig}
\usepackage{color}
\usepackage{appendix}

\begin{document}

\title{Compressibility and spin susceptibility \\ 
in the evolution from BCS to BEC superfluids}

\author{Kangjun Seo and Carlos A. R. S{\'a} de Melo}
\affiliation{School of Physics, Georgia Institute of Technology, 
Atlanta, Georgia 30332, USA}
\date{\today}

\begin{abstract}
We describe the relation between the isothermal atomic compressibility and density fluctuations in mixtures of two-component fermions with population or mass imbalance. We derive a generalized version of the fluctuation-dissipation theorem which is valid for both balanced and imbalanced Fermi-Fermi mixtures. Furthermore, we show that critical exponents for the compressibility can be extracted via an analysis of the density fluctuations when phase transitions occur as a function of population imbalance or interaction parameter.
In addition, we demonstrate that in the presence of trapping potentials, the local compressibility, local spin-susceptibility and local density-density correlations can be extracted from experimental data via a generalized local fluctuation-dissipation theorem which is valid beyond the local density approximation. 
Lastly, we use the local density approximation to calculate theoretically the local compressibility, local spin-susceptibility and local density-density fluctuations to compare with experimental results as they become available.  
\end{abstract}

\maketitle

\section{Introduction}
\label{sec:introduction}
Very recently experimental advances in Bose and Fermi systems have allowed for studies of density fluctuations and the use of the fluctuation dissipation theorem to obtain information about some thermodynamic properties of ultra-cold atoms. 
In the fermion case, the measurement of density fluctuations and of the atomic compressibility was performed for non-interacting three-dimensional systems in harmonic traps~\cite{ketterle-10, esslinger-10}, while in the boson case, the connection between density fluctuations and compressibility was used to study superfluidity in a two-dimensional system, and extract critical exponents associated with the transition from a normal Bose gas to a Berezinskii-Kosterlitz-Thouless superfluid~\cite{chin-10}. The experimental extraction of the isothermal compressibility from measurements of density fluctuations was suggested several years ago both in harmonically confined systems~\cite{iskin-05} and optical lattices~\cite{iskin-06a}, but only recently improvements in the detection schemes of density fluctuations became sufficiently sensitive to extract this information from experimental data~\cite{raizen-05,bloch-05,bouchoule-06,steinhauer-10,stringari-11}.

We see no major technical impediment to use techniques that are sensitive to density fluctuations in population imbalanced Fermi-Fermi mixtures with equal masses~\cite{ketterle-06, hulet-06} or with unequal masses~\cite{grimm-10,grimm-11}, where, in principle, the compressibility and spin susceptibility matrix elements can be directly extracted from the density and density fluctuation profiles.  
In fact, in a very recent experiment~\cite{ketterle-11} using laser speckles, the isothermal compressibility and the spin susceptibility were measured as a function of interaction parameter via the fluctuation-dissipation theorem throughout the evolution from BCS to BEC superfluidity in 
balanced Fermi systems.

In this paper, we extend some of our recent results~\cite{seo-11}, and show that through local measurements of densities and density fluctuations several local and global thermodynamic properties can be extracted.
We derive a generalized version of the fluctuation-dissipation theorem valid for any mixtures of atoms, and use it to analyze density fluctuations and the compressibility of mixtures of two-component (or two types of) fermions with and without population imbalance at low temperatures.  
For spatially uniform systems, we show that the global compressibility can be extracted from measurements of the density and density fluctuations for each component, while for spatially non-uniform systems, we show that the local and global compressibility can be extracted from measurements of the local density and local density fluctuations for each component.
Finally, we also show that across the boundaries between different phases critical exponents can be extracted provided that there is sufficiently spatial resolution for the critical region to be experimentally accessed.

This paper is organized as follows.
In section~\ref{sec:hamiltonian}, we describe the hamiltonian corresponding to balanced or imbalanced Fermi mixtures of equal or unequal masses.
In section~\ref{sec:compressibility}, we analyze the compressibility matrix, and derive a generalized fluctuation dissipation theorem relating the compressibility matrix and density fluctuations for balanced or imbalanced two-component Fermi system. Furthermore, we establish a connection  
between the isothermal compressibility or the spin susceptibility to the compressibility matrix elements.  
In section~\ref{sec:compressibility-trap}, we discuss the effects of traps and derive a local generalized fluctuation-dissipation theorem which is valid beyond the local density approximation (LDA). Using this theorem, we describe the local isothermal compressibility and the local spin-susceptibility, and their relation to the local compressibility matrix.
In Section~\ref{sec:fermi-fermi-mixtures}, we apply the results obtained in the previous sections to describe Fermi-Fermi mixtures of equal or unequal masses, as well as equal or unequal populations without a trap.
In particular, the compressibility and spin-susceptibility are calculated as a function of population imbalance and interaction parameter for the 
case of mixtures of equal mass fermions.
In Section~\ref{sec:fermi-fermi-mixtures-trap}, we use the local fluctuation-dissipation theorem and the local density approximation to calculate 
the spatial dependence of the local compressibility matrix, local isothermal compressibility and local spin-susceptibility for population balanced and imbalanced mixtures of fermions with equal masses. 
In section~\ref{sec:summary-conclusions}, we summarize our main findings and provide some concluding remarks.

\section{Hamiltonian}
\label{sec:hamiltonian}
To investigate the physics described above, we start with the real space Hamiltonian ($\hbar = 1$)
\begin{equation}
\label{eqn:hamiltonian-real-space}
\hat H = \hat H_{0} + \hat H_{\rm trap} +  \hat H_{\rm int}
\end{equation} 
for a three dimensional $s$-wave superfluid.
Here, $\hat H_{0}$ describes the kinetic energy of fermions 
\begin{equation}
\hat H_{0} 
=
\sum_{\alpha} \int d {\bf r}\,
\hat \psi^{\dagger}_\alpha ({\bf r})
\left[ 
- 
\frac{\nabla^2}{2m_\alpha}
-\mu_\alpha 
\right]
\hat \psi_\alpha ({\bf r}),
\end{equation}
where $\hat \psi_{\alpha}^{\dagger} ({\bf r})$ represents the creation of fermions of species ${\alpha}$ with mass $m_{\alpha}$, and $\mu_{\alpha}$ is the chemical potential which determines the average number of fermions for each species $\alpha$.
The term that contains the trapping potential is 
\begin{equation}
\label{eqn:hamiltonian-trap}
\hat H_{\rm trap} 
=
\sum_{\alpha} \int d {\bf r} \,
V_\alpha ( {\bf r} ) \,
\hat n_\alpha ( {\bf r} ),
\end{equation}
where $
\hat n_\alpha ( {\bf r}) 
=
\hat \psi_\alpha^{\dagger} ( {\bf r} )
\hat \psi_\alpha ( {\bf r} )
$ is the particle density operator of fermions of species $\alpha$, and $V_\alpha ({\bf r})$ is the corresponding trapping potential which 
may vary from one species to another.
Since we are confining ourselves to $s$-wave interactions the interaction term is
\begin{equation}
\hat H_{\rm int}
=
\int d {\bf r}\, d {\bf r^\prime}\,
V_{\rm int} ( {\bf r}, {\bf r}^{\prime} ) \,
\hat n_\uparrow ( {\bf r} ) \,
\hat n_\downarrow ( {\bf r^\prime} ),
\nonumber
\end{equation}
involving only spin-up and spin-down densities. 
In addition, we consider the case of an attractive point contact interaction $
V_{\rm int} ({\bf r},{\bf r}^\prime)
=
-g \delta ( {\bf r} - {\bf r}^\prime )
$ with $g>0$.

By allowing each fermion type to have their own chemical potential $\mu_{\alpha}$ and mass $m_\alpha$, we can investigate thermodynamic properties of population imbalanced mixtures of Fermi systems with equal masses $m_{\uparrow} = m_{\downarrow} = m$, and more generally, unequal masses $m_{\uparrow} \ne m_{\downarrow}$.
From the grand partition function 
\begin{equation}
\label{eqn:grand-partition-function}
\mathcal Z = {\rm Tr}  \exp \left(-\hat H/T \right) ,
\end{equation} 
we can define the thermodynamic potential 
\begin{equation}
\label{eqn:omega-potential}
\Omega 
= 
- T \ln \mathcal Z,
\end{equation}
and obtain global thermodynamic properties such as the average number of particles, entropy, specific heat, compressibility, and so on. 
In Fermi-Fermi mixtures, the compressibility is an important global thermodynamic property which can now be directly measured experimentally in the context of ultra-cold atoms, thus we discuss it next.

\section{Compressibility matrix without trapping potential}
\label{sec:compressibility}
In this section, we discuss the compressibility matrix, the isothermal compressibility, and the spin susceptibility for population balanced and imbalanced Fermi-Fermi mixtures in two situations.
First, we ignore the trapping potential $V_{\alpha} ({\bf r})$, and analyze the spatially homogeneous system to simplify the discussion.
After that, we add back the trapping potential $V_{\alpha} ({\bf r})$, and return to the spatially inhomogeneous system, which is more relevant to experiments, as described in Sec.~\ref{sec:compressibility-trap}.

\subsection{Compressibility Matrix} 
\label{sec:compressibility-no-trap}
Setting the trapping potential $V_{\alpha} ({\bf r})$ to zero in Eq.~(\ref{eqn:hamiltonian-trap}) leads to no-trap Hamiltonian
\begin{equation}
\label{eqn:hamiltonian-no-trap}
\hat H 
=
\hat H_1 - \sum_{\alpha} \mu_{\alpha} {\hat N}_{\alpha},
\end{equation}
where the first term is given by 
\begin{equation}
\label{eqn:hamiltonian-1}
\hat H_1 
= 
\sum_{\alpha}
\int 
d{\bf r}
\hat \psi^{\dagger}_{\alpha} ({\bf r}) 
\left[
-
\frac{\nabla^2}{ 2m_{\alpha} } 
\right]
\hat \psi_\alpha ({\bf r})
+ 
\hat H_{\rm int},
\end{equation}
representing the sum of the kinetic energy and interaction contributions, and $
{\hat N}_{\alpha}
=
\int 
d{\bf r} \,
\hat n_\alpha ({\bf r})
$ the number operator for hyperfine state $\alpha$.
With this explicit separation, it becomes easier to see that the average number of particles $N_{\alpha} = \langle {\hat N_\alpha} \rangle$, 
defined by the thermodynamic average
\begin{equation}
\langle {\hat N_\alpha} \rangle 
= 
\mathcal Z^{-1} 
{\rm Tr} 
\left[ {\hat N_\alpha} \,
e^{-(\hat H_1 - \sum_{\gamma} 
\mu_{\gamma} \hat N_{\gamma} )/T}
\right],
\end{equation}
can be rewritten in terms of the thermodynamic potential $\Omega$ as
\begin{equation}
\label{eqn:number-alpha}
N_{\alpha} = 
\langle \hat N_{\alpha} \rangle 
= 
- 
\left( 
\frac{
\partial \, \Omega
}{
\partial \mu_{\alpha}
} 
\right)_T
\end{equation}

Starting from Eq.~(\ref{eqn:number-alpha}), we define the pseudo-compressibility matrix $\{ \tilde \kappa \}$ through its matrix elements
\begin{equation}
\label{eqn:pseudo-compressibility-1}
{\tilde \kappa}_{\alpha \beta} 
=
T \left(
\frac{
\partial N_{\alpha} 
}{
\partial \mu_{\beta} 
} \right)_T
=
- T \left(
\frac{
\partial^2 \, \Omega 
}{
\partial \mu_{\alpha} \partial \mu_{\beta}
} \right)_T.
\end{equation}
Using the fact that ${\hat N}_{\alpha}$ commutes with $\hat H_1$ defined in Eq.~(\ref{eqn:hamiltonian-1}) and using the definition of $\Omega$ in Eq.~(\ref{eqn:omega-potential}), we can calculate the matrix elements of $\{\tilde \kappa\}$ explicitly, leading to 
\begin{equation}
\label{eqn:pseudo-compressibility-2}
{\tilde \kappa}_{\alpha \beta} 
=
\langle 
{\hat N}_{\alpha} 
{\hat N}_{\beta} 
\rangle
-
\langle 
{\hat N}_{\alpha} 
\rangle
\langle
{\hat N}_{\beta} 
\rangle,
\end{equation}
expressed in terms of thermodynamic averages.
Here, the correlated average of number operators ${\hat N}_{\alpha}$ and ${\hat N}_{\beta}$ is 
\begin{equation}
\langle 
{\hat N}_{\alpha} 
{\hat N}_{\beta} 
\rangle
=
\mathcal Z^{-1} 
{\rm Tr} 
\left[ 
{\hat N}_{\alpha} {\hat N}_{\beta} \,
e^{-\hat H/T}
\right].
\end{equation}

The mechanical stability of the system at any finite temperature $T$ is guaranteed when both eigenvalues of the matrix $\{ \tilde \kappa \}$ are positive definite, while in the limit of $T \to 0$ it is the positivity of both eigenvalues of $ {\lim}_{T \to 0} ( \{ \tilde \kappa \}/T ) $ that guarantees the stability of the system.
At finite $T$, the eigenvalues of $\{ \tilde \kappa \}$ are positive if the determinant ${\rm det}\left[\, \{\tilde \kappa\} \,\right] > 0$ and the trace ${\rm Tr} \left[\, \{\tilde \kappa\} \,\right] > 0$. 
Analogously, at $T \to 0$, the eigenvalues of $\lim_{T \to 0} (\{\tilde \kappa\}/T)$ are positive if the determinant ${\rm det} \left[\,\lim_{T \to 0} (\{\tilde \kappa\}/T)\,\right] > 0$ and the trace ${\rm Tr} \left[\,\lim_{T \to 0} (\{\tilde \kappa\}/T)\,\right] > 0$.

Using the definition $
N_{\alpha} 
= 
\langle
{\hat N}_{\alpha}
\rangle,
$ we can  explicitly identify that the pseudo-compressibility matrix is a measure of the correlation in density-density (or number-number) fluctuations, since it can be rewritten as 
\begin{equation}
\label{eqn:fluctuation-dissipation}
{\tilde \kappa}_{\alpha\beta} 
= 
\langle
\Delta \hat N_{\alpha} \,\Delta \hat N_{\beta}\rangle,
\end{equation}
where $\Delta \hat N_{\alpha}
=
\hat N_{\alpha} - N_{\alpha}.$
The corresponding generalized compressibility matrix $\{\kappa\}$ can be obtained from $\{\tilde \kappa\}$ through the relation 
\begin{equation}
\label{eqn:compressibility-matrix}
\kappa_{\alpha \beta} 
= 
\tilde \kappa_{\alpha \beta} /( N_\alpha N_\beta ),
\end{equation}
describing a generalized fluctuation-dissipation theorem for multicomponent fermions, which is valid for fermions of equal or unequal masses and equal or unequal populations.
As it is discussed below, when the populations for each fermion species are the same $N_{\uparrow} = N_{\downarrow}$, this more general result reduces to the standard form of the fluctuation-dissipation theorem~\cite{kubo-57}.

In order to be more explicit, we note that the compressibility matrix element $\kappa_{\uparrow \uparrow}$ is directly related to the particle number fluctuation in the spin-up/spin-up channel $
\langle ( \Delta \hat N_\uparrow )^2 \rangle 
$, that $\kappa_{\downarrow \downarrow}$ is directly related to the particle number fluctuation in the spin-down/spin-down channel $
\langle (\Delta\hat N_\downarrow)^2 \rangle 
$, and that $\kappa_{\uparrow \downarrow}$ is directly related to the correlation between the particle number fluctuation in the spin-up/spin-down channel $
\langle 
\Delta \hat N_\uparrow \Delta \hat N_\downarrow 
\rangle 
$.
In addition, the matrix $\{\kappa\}$ is symmetric, such that $
\kappa_{\uparrow \downarrow} 
= 
\kappa_{\downarrow \uparrow}.
$

In very recent experiments~\cite{ketterle-10,esslinger-10}, particle number (density) fluctuations were directly observed for an ideal (non-interacting) Fermi gas of equal masses and equal populations $
N_\uparrow = N_\downarrow,
$ in which case the chemical potentials $\mu_\alpha$ for each hyperfine state $\alpha$ are exactly the same $(\mu_\uparrow = \mu_\downarrow)$.
For arbitrary interactions, identical masses and populations, it is sufficient to analyze fluctuations only in the total average number of particles $ N = N_\uparrow + N_\downarrow $ to extract the isothermal compressibility 
\begin{equation}
\label{eqn:compressibility-balanced}
\kappa_T 
=
- \frac{1}{V}
\left(
\frac{
\partial V
}{
\partial P
} 
\right)_T
= 
\frac{V}{N^2}
\left(
\frac{
\partial N
}{
\partial \mu
} 
\right)_T. 
\end{equation}
Using the standard form of the fluctuation-dissipation theorem for a balanced system, the isothermal compressibility can be rewritten as 
\begin{equation}
\label{eqn:compressibility-balanced-fluctuation}
\kappa_T 
= 
\frac{V}{T}
\left(
\frac{
\langle {\hat N}^2 \rangle - N^2 
}{
N^2
}
\right).
\end{equation}

For mass or population imbalanced systems, however, the relation between the isothermal compressibility $
\kappa_T = - (\partial V /\partial P)_T / V
$ and the fluctuations in the number of particles is more subtle.
The determination of $\kappa_T$ involves an analysis of the pseudo-compressibility matrix $\{\tilde \kappa\}$, which contains all the information  
about fluctuations in particle numbers (densities) for each hyperfine state. 
Using similar experimental techniques as those described in Refs.~\cite{ketterle-10,esslinger-10}, it should be possible to measure the matrix elements of $\{\tilde \kappa\}$ directly not only for the ideal non-interacting Fermi system, but also when strong correlations are present, as the fluctuation dissipation theorem described above is valid for all interaction strengths.
Thus, it is important to identify the relation between the isothermal thermodynamic compressibility $\kappa_T$ and the elements of the compressibility matrix $\{\kappa\}$. 
This connection is discussed next.

\subsection{Isothermal compressibility}
\label{sec:isothermal compressibility}
The relation between $\kappa_T$ and $\kappa_{\alpha \beta}$ can be established via the thermodynamic potential $\Omega = -PV$, where $P$ is the pressure and $V$ is the volume of the system. 
Defining $G = \Omega + PV = 0$, and recalling that $\Omega$ is a function of temperature $T$, volume $V$ and chemical potentials $\mu_\uparrow$, $\mu_\downarrow$ results in
\begin{equation}
\label{eqn:gibbs-free-energy}
dG 
= 
-SdT 
+ 
V dP 
- \sum_{\beta = \uparrow,\downarrow} 
N_\beta d \mu_\beta
= 
0.
\end{equation}
At constant temperature $dT = 0$, we can establish the relation 
\begin{equation}
\label{eqn:VdP}
V dP\vert_T 
= 
\sum_\alpha
N_\beta d \mu_\beta \vert_{T}.
\end{equation}
This means that the inverse isothermal compressibility (also called the Bulk modulus)
\begin{equation}
\frac{1}
{\kappa_T} 
= 
- V 
\left(
\frac{
\partial P
}{
\partial V
} 
\right)_T
\end{equation}
can be written in terms of isothermal partial derivatives of $\mu_{\beta}$ with respect to volume
\begin{equation}
\frac{1}{\kappa_T}
= 
- \sum_{\beta}
N_\beta 
\left(
\frac{
\partial \mu_\beta 
}{ 
\partial V 
}
\right)_{T,N_\uparrow,N_\downarrow}.
\end{equation}
But, in turn, the partial derivatives $(\partial \mu_{\beta} / \partial V )_T$ can be expressed in the terms of the average numbers $N_\uparrow$ and $N_\downarrow$, since $\mu_\beta  = \mu_\beta ( N_\uparrow/V, N_\downarrow/V)$. For the isothermal variation of the spin-$\beta$ chemical potential with respect to the volume $V$, we can write 
\begin{equation}
-V 
\left(
\frac{
\partial \mu_{\beta} 
}{
\partial V 
} 
\right)_{T,N_\uparrow,N_\downarrow}
= 
\sum_\alpha
N_\alpha
\left(
\frac{
\partial \,\mu_{\beta}
}{
\partial N_\alpha
} 
\right)_{T} .
\end{equation}
The last step in establishing the relationship between $\kappa_T$ and $\tilde \kappa_{\alpha \beta}$, just requires the use of the definition $ 
( \partial N_\alpha / \partial \mu_\beta )_T 
= 
\tilde \kappa_{\alpha \beta} / T
$ from Eq.~(\ref{eqn:pseudo-compressibility-1}) or its inverse $ 
(\partial \mu_\beta / \partial N_\alpha )_T 
=
T 
/
\tilde \kappa_{\alpha \beta}. 
$ 
Putting all this together leads to the interesting relation
\begin{equation}
\label{eqn:isothermal-compressibility}
\frac{1}{\kappa_T}
=
\frac{T}{V}
\left(
\frac{N_\uparrow^2}{\tilde \kappa_{\uparrow \uparrow}}
+
\frac{N_\uparrow N_\downarrow}{\tilde \kappa_{\uparrow \downarrow}}
+
\frac{N_\downarrow N_\uparrow}{\tilde \kappa_{\downarrow \uparrow}}
+
\frac{N_\downarrow^2}{\tilde \kappa_{\downarrow \downarrow}}
\right),
\end{equation}
which can still be further written in a compact form 
\begin{equation}
\label{eqn:isothermal-compressibility-compact}
\frac{1}{\kappa_T}
= 
\frac {T}{V} 
\sum_{\alpha \beta} 
\left( 
\frac{1}{\kappa_{\alpha \beta}} 
\right),
\end{equation}
where we used the definition of the compressibility matrix elements $\kappa_{\alpha \beta} = \tilde \kappa_{\alpha \beta}/(N_\alpha N_\beta)$ described in Eq.~(\ref{eqn:compressibility-matrix}).
From the relation between the pseudo-compressibility matrix elements $\tilde \kappa_{\alpha \beta}$ and the density-density (number-number) fluctuations shown in Eq.~(\ref{eqn:fluctuation-dissipation}), we can rewrite the inverse of the isothermal compressibility as
\begin{equation}
\label{eqn:isothermal-compressibility-compact1}
\frac{1}{\kappa_T} =
\frac{T}{V} \sum_{\alpha\beta}
\left( 
\frac{
N_{\alpha} N_{\beta}
}{
\langle \Delta N_{\alpha} \, \Delta N_{\beta}\rangle} 
\right).
\end{equation}

The result given in Eq.~(\ref{eqn:isothermal-compressibility-compact}) reminds us of the formula for the computation of the equivalent resistance  of four different resistors in parallel. Each matrix element, $1 / \kappa_{\alpha \beta},$ acts as independent parallel channel of number fluctuations, which combined give essentially the inverse isothermal compressibility. Measurements of the compressibility matrix elements $\kappa_{\alpha \beta}$ are important as they can be connected further to the pseudo-spin susceptibility of the system, as discussed next.

\subsection{Connection to pseudo-spin susceptibility}
\label{sec:spin-susceptibility}
At this point it is important to mention that even though we have chosen to work with particle numbers $N_\uparrow$, $N_\downarrow$, and chemical  potentials $\mu_\uparrow$, $\mu_\downarrow$, we could have chosen also to work with the total number of particles 
\begin{equation}
N_+ 
=  
N_\uparrow 
+ 
N_\downarrow,
\end{equation}
the difference in particle numbers
\begin{equation}
N_- 
= 
N_\uparrow 
- 
N_\downarrow,
\end{equation}
and their corresponding chemical potentials $\mu_\pm = ( \mu_\uparrow \pm \mu_\downarrow )/2 $, respectively.
This new choice provides the same information about the system since the term involving the chemical potentials in the Hamiltonian $H$ of Eq.~(\ref{eqn:hamiltonian-no-trap}) can be rewritten as
\begin{equation}
\sum_{\alpha} \mu_\alpha \hat N_\alpha
= 
\mu_+ \hat N_+ + \mu_- \hat N_-,
\end{equation}
such that the corresponding partition function $\mathcal Z$ or thermodynamic potential $\Omega$ also contain the same terms. 
In this case, we can define a similar pseudo-compressibility matrix 
\begin{equation}
{\tilde \kappa}_{ij} 
=
T 
\left(
\frac{
\partial N_i 
}{
\partial \mu_j
} 
\right)_T
=
- 
T 
\left(
\frac{
\partial^2 \, \Omega
}{
\partial \mu_{i} \partial \mu_{j}
} 
\right)_T, 
\end{equation}
where each of the indices $i,j$ can take $\pm$ values.
An explicit calculation using the grand partition function $\mathcal Z$ allows to express ${\tilde \kappa}_{ij}$ in terms of the thermodynamic averages
\begin{equation}
{\tilde \kappa}_{ij} 
=
\langle 
{\hat N}_i 
{\hat N}_j
\rangle
-
\langle 
{\hat N}_i 
\rangle
\langle
{\hat N}_j 
\rangle.
\end{equation}
The connection between the pseudo-compressibility matrices in the two different representations is simple.
The first diagonal term is $
{\tilde \kappa}_{++} = 
{\tilde \kappa}_{\uparrow \uparrow}
+
{\tilde \kappa}_{\downarrow \downarrow}
+
2{\tilde \kappa}_{\uparrow \downarrow},
$ the second diagonal term is $
{\tilde \kappa}_{--} = 
{\tilde \kappa}_{\uparrow \uparrow}
+
{\tilde \kappa}_{\downarrow \downarrow}
-
2{\tilde \kappa}_{\uparrow \downarrow},
$ while the off-diagonal terms $
{\tilde \kappa}_{+-} 
=
{\tilde \kappa}_{-+}
=
{\tilde \kappa}_{\uparrow \uparrow}
-
{\tilde \kappa}_{\downarrow \downarrow}
$ are identical since the matrix ${\tilde \kappa}_{ij}$ is symmetric.

The corresponding expression of the isothermal compressibility $\kappa_T$ has exactly the same form as before, just with indices $(\uparrow, \downarrow)$ mapped into indices $(+, -)$, and thus leading to 
\begin{equation}
\label{eqn:isothermal-compressibility-notrap-pm}
\frac{1}{\kappa_T} 
=
\frac{T}{V}
\left(
\frac{N_+^2}{\tilde \kappa_{+ +}}
+
\frac{N_+ N_-}{\tilde \kappa_{+ -}}
+
\frac{N_-  N_+}{\tilde \kappa_{- +}}
+
\frac{N_-^2}{\tilde \kappa_{- -}}
\right).
\end{equation}
This way of writing $\kappa_T$ is more transparent because the population balanced case $N_\uparrow = N_\downarrow$ corresponds to $N_-  = 0$, and thus leads to the standard expression for the compressibility 
\begin{equation}
\frac{1}{\kappa_T} 
= 
\frac{T}{V} 
\left(
\frac{
N_+^2
}{
\tilde \kappa_{++}
}
\right), 
\end{equation}
which upon inversion and further identification of $N_+ \to N$ leads to the balanced form of the fluctuation dissipation theorem as described 
in Eq.~(\ref{eqn:compressibility-balanced-fluctuation}).

For mass or population imbalanced cases, the direct use of Eq.~(\ref{eqn:isothermal-compressibility-notrap-pm}) allows for the extraction of the atomic compressibility directly from measurements of densities and density-density fluctuations. 
This result generalizes initial suggestions that the isothermal compressibility could be measured experimentally for balanced systems in optical lattices or harmonic traps~\cite{iskin-05, iskin-06a} via the fluctuation dissipation theorem.

Notice that the dimensionless pseudo-spin susceptibility $\chi_{zz}$ of the system can also be extracted from the pseudo-compressibility matrix $\tilde \kappa_{ij}$, since $\mu_{-} = (\mu_\uparrow - \mu_\downarrow)/2$ plays the role of an effective magnetic field $h_z$ along the quantization axis $z$, and $N_{-}=N_\uparrow - N_\downarrow$ plays the role of the magnetization $m_z$. For imbalanced Fermi systems, this implies that 
\begin{equation}
\label{eqn:pseudo-spin-susceptibility}
\tilde \kappa_{--} 
=  
T 
\left(
\frac{ 
\partial  N_{-} 
}{
\partial \,\mu_{-} 
} 
\right)_T
=
T
\left(
\frac{ 
\partial m_z 
}{ 
\partial \, h_z 
} 
\right)_T
=
T \, \chi_{zz},
\end{equation}
and generalizes the results obtained for ultra-cold fermions interacting via $p$-wave interactions (triplet channel)~\cite{botelho-05}.

Now that the case of zero trapping potential has been analyzed, we add back $V_{\alpha} ({\bf r})$ to the Hamiltonian of the system and discuss its effects next.

\section{Compressibility matrix in a trap}
\label{sec:compressibility-trap}
To consider the most general case, where the external trapping potential $V_{\alpha} ({\bf r})$ is non-zero, we include a source term in the Hamiltonian defined in Eq.~(\ref{eqn:hamiltonian-real-space}), leading to
\begin{equation}
\hat H_J = \hat H - \sum_{\alpha} \int d{\bf r} 
J_\alpha({\bf r}) \,
\hat n_\alpha ({\bf r}) .
\end{equation}
In the presence of this source term, the grand partition function $\mathcal Z$ becomes a generating functional of the $J_\alpha$ as well as temperature $T$, volume $V$, and the chemical potential $\mu_\alpha$,
\begin{equation}
\mathcal Z\,[\, T,V,\mu_\alpha,J_\alpha({\bf r}) \,]
= 
{\rm Tr}\,
e^{- H_J /T}.
\end{equation}
The thermodynamic potential $$
\Omega 
= 
-T 
\ln \mathcal Z\,[\,T,V,\mu_\alpha,J_\alpha({\bf r})\,]
$$ is also a functional of $J_\alpha({\bf r})$, such that local and even non-local thermodynamic quantities can be obtained. 
For instance, the local density $
n_\alpha ({\bf r}) 
= 
\langle \hat n_\alpha({\bf r})\rangle$ is simply written as the functional derivative
\begin{equation}
n_\alpha({\bf r})
=
-T \, 
\frac{
\delta\,\Omega 
}{
\delta J_\alpha({\bf r})
} 
\Big\vert_{J_\uparrow,J_\downarrow \to 0}.
\end{equation}
And the second derivative gives the correlation of the particle density fluctuation
\begin{equation}
\langle 
\Delta {\hat n}_{\alpha} ({\bf r}) \,
\Delta {\hat n}_{\beta} ({\bf r}^\prime) 
\rangle
=
-T \, 
\frac{
\delta^2 \,\Omega 
}{
\delta J_\alpha({\bf r})\, \delta J_\beta({\bf r}^\prime) 
}
\Big\vert_{J_\uparrow,J_\downarrow \to 0},
\end{equation}
with $
\Delta {\hat n}_{\alpha} ({\bf r}) 
= 
{\hat n}_{\alpha} ({\bf r})
-
n_{\alpha} ({\bf r}).$
Correspondingly, we define the non-local pseudo-compressibility matrix elements $\tilde \kappa_{\alpha \beta}$ as
\begin{equation}
\tilde \kappa_{\alpha \beta} ({\bf r}, {\bf r}^\prime)
= 
\langle 
\Delta {\hat n}_{\alpha} ({\bf r}) \,
\Delta {\hat n}_{\beta} ({\bf r}^\prime) 
\rangle.
\end{equation}
Then, the local pseudo-compressibility matrix is simply defined as $
\tilde \kappa_{\alpha \beta} ({\bf r}) 
=
{\tilde \kappa}_{\alpha \beta} ( {\bf r}, {\bf r} ).
$

The inverse of the local isothermal compressibility $\kappa_T ({\bf r})$ can be expressed in terms of the local pressure density $ p({\bf r}) $ as 
\begin{equation}
\frac{1}{\kappa_T ({\bf r})} 
= 
- V 
\left(
\frac{ 
\partial p ({\bf r})
}{
\partial \, V
} 
\right)_T, 
\end{equation}
where the total pressure is $P = \int d{\bf r}\, p({\bf r})$.
Following the same steps used to derive Eq.~(\ref{eqn:isothermal-compressibility}), we arrive to the local relation 
\begin{equation}
\frac{1}{ \kappa_T ({\bf r}) }
= 
\frac{T}{V}
\sum_{\alpha \beta} 
\frac{1}{\kappa_{\alpha \beta} ({\bf r})},
\end{equation}
where the local compressibility matrix becomes
\begin{equation}
\kappa_{\alpha \beta} ({\bf r}) 
=
\frac{
{\tilde \kappa}_{\alpha \beta} ({\bf r})
}{
n_{\alpha} ({\bf r}) \, n_{\beta} ({\bf r})  
},
\end{equation}
thus revealing a local generalization of the fluctuation-dissipation theorem.

Analogously to the zero trap case, instead of working with the local particle number densities $n_\alpha ({\bf r})$, we could have chosen to work  with the total number of particles 
\begin{equation}
n_+ ({\bf r}) 
=  
n_\uparrow ({\bf r})
+ 
n_\downarrow ({\bf r}),
\end{equation}
and the difference in the number of particles 
\begin{equation}
n_-({\bf r})
= 
n_\uparrow ({\bf r})
- 
n_\downarrow ({\bf r}).
\end{equation}
In this new basis, we define a similar pseudo-compressibility matrix 
\begin{equation}
\tilde \kappa_{ij} ({\bf r}, {\bf r}^\prime)
= 
\langle 
\Delta {\hat n}_{i} ({\bf r}) \,
\Delta {\hat n}_{j} ({\bf r}^\prime) \rangle.
\end{equation}

The corresponding expression for the local isothermal compressibility $\kappa_T ({\bf r})$ has exactly the same form as before, just with indices  
$(\uparrow, \downarrow)$ mapped into indices $(+, -)$, thus leading to 
\begin{equation}
\label{eqn:isothermal-compressibility-pm}
\frac{1}{\kappa_T ({\bf r})}
=
\frac{T}{V}
\sum_{ij}
\frac{1}{\kappa_{ij}({\bf r})},
\end{equation}
where the local compressibility matrix elements are  
\begin{equation}
\kappa_{ij} ({\bf r})
=
\frac{
\tilde \kappa_{ij}({\bf r})
}{ 
n_i ({\bf r}) \, n_j ({\bf r}) 
}.
\end{equation}
The use of equation Eq.~(\ref{eqn:isothermal-compressibility-pm}) allows for the extraction of the local isothermal compressibility $\kappa_T ({\bf r})$ and the local dimensionless pseudo-spin susceptibility 
\begin{equation}
\chi_{zz}({\bf r}) 
= 
\frac{\tilde \kappa_{--}({\bf r})}{T} 
= 
\frac{\langle (\Delta \hat n_{-}({\bf r}))^2 \rangle}{T}
\end{equation}
directly from the measurements of the local densities and local density fluctuations for imbalanced systems.

Now that the case of non-zero trapping potential has been analyzed, we turn our attention to the application of these results to the cases of 
balanced and imbalanced Fermi-Fermi mixtures.

\section{Fermi-Fermi mixtures without a trap}
\label{sec:fermi-fermi-mixtures}
So long as the particle number operators commute with the Hamiltonian of the system, the results discussed in section~\ref{sec:compressibility} apply to any binary mixtures (Bose-Bose, Fermi-Bose, Fermi-Fermi) in the case where there is no trap; while the results of section~\ref{sec:compressibility-trap} apply to any binary mixtures in a trap. 
In this section, we discuss generally the cases of Fermi-Fermi mixtures of equal or unequal masses when populations are balanced or imbalanced, but we focus particularly in the case of equal mass mixtures with balanced or imbalanced populations as a concrete example.

\subsection{Thermodynamic Potential without a Trap}
\label{sec:fermi-fermi-thermodynamic-potential}
As an specific example of the general relations just derived, we discuss the case of imbalanced Fermi-Fermi mixtures with equal masses, which has  attracted substantial interest~\cite{yip-06, mueller-06, iskin-06b, stoof-06, pieri-06, tempere-07} recently.

The action corresponding to the Hamiltonian given in Eq.~(\ref{eqn:hamiltonian-real-space}) is simply written in the coherent state representation as
\begin{equation}
\label{eqn:fermi-fermi-action}
S 
=
\int_0^\beta d\tau
\int d{\bf r} 
\left[
\bar{\psi}_{\alpha} ({\bf r}, \tau)
\partial_{\tau}
\psi_{\alpha} ({\bf r}, \tau)
+
{\cal H} ({\bf r}, \tau)
\right],
\end{equation}
where $\bar \psi_{\alpha} ( {\bf r}, \tau)$, and $\psi_{\alpha} ({\bf r}, \tau)$ are Grassman variables for the fermion type $\alpha$.
The corresponding partition function is then written as the functional integral
\begin{equation}
\mathcal Z 
=
\int 
D
\left[
\bar \psi_{\alpha}
\psi_{\alpha} 
\right]
\exp 
\left(
- 
S
\left[
\bar \psi_{\alpha}, 
\psi_{\alpha} 
\right]
\right),
\end{equation}
from which the thermodynamic potential $\Omega = - T \ln \mathcal Z$ can be directly calculated.

Since we are considering first the case without a trapping potential, $V_{\alpha} ({\bf r}) = 0$, we can use the translational invariance of the system to write the Hamiltonian $H(\tau) = \int d{\bf r} \,\mathcal{H}({\bf r},\tau)$ as
\begin{equation}
H(\tau)
=
\sum_{{\bf k}, \alpha}
\xi_{{\bf k}, \alpha} 
\bar \psi_{{\bf k},\alpha}(\tau)  \,
\psi_{{\bf k},\alpha} (\tau)
-
g \sum_{\bf q}
\bar B_{\bf q} (\tau) B_{\bf q} (\tau)
\end{equation}
where $
\xi_{{\bf k}, \alpha} 
= 
\epsilon_{{\bf k},\alpha}
- 
\mu_{\alpha},
$ with $
\epsilon_{{\bf k},\alpha} = k^2/(2m_\alpha)$ being the single particle dispersion. Here, $\psi_{{\bf k},\alpha}(\tau)$ is the Fourier transform of $\psi_\alpha({\bf r},\tau)$, and 
\begin{equation}
B_{\bf q} (\tau) = \sum_{\bf k} \Gamma_{\bf k} \,
\psi_{{\bf q}/2-{\bf k}, \downarrow} (\tau) \,
\psi_{{\bf q}/2+{\bf k}, \uparrow} (\tau)
\end{equation}
describes paired fermions and $\Gamma_{\bf k}$ corresponds to the symmetry of the pairing interaction. For instance, in the case of $s$-wave pairing we have $\Gamma_{\bf k} = 1$.

The effective action $S$ has been successfully calculated for the case of a uniform superfluid~\cite{iskin-07} in the Gaussian approximation 
as
\begin{equation}
S_{G} 
= 
S_0 
+ 
\frac{1}{2T}
\sum_{q} 
\bar \Lambda (q)
{\bf F}^{-1}(q)
\Lambda (q),
\nonumber
\end{equation}
where $q = ({\bf q}, \nu_{\ell} )$, with $\nu_{\ell} = 2\pi \ell T$ being the bosonic Matsubara frequency at temperature $T$.
Here, $\Lambda (q)$ is the order parameter fluctuation field and the matrix ${\bf F}^{-1} (q)$ is the inverse fluctuation propagator.
The explicit expression of ${\bf F}^{-1} (q)$ is given in Appendix~\ref{app-a}.

The saddle point action $S_0$ is given by
\begin{equation}
S_0 
=
\frac{\vert \Delta_0 \vert^2}{g T}
+
\frac{1}{T}
\sum_{\bf k} 
\left(
\xi_{ {\bf k}, + } - \xi_{ {\bf k}, - }
\right)
+ 
\sum_{\nu=1,2}
\ln 
\left[
n_F 
(
- \mathcal E_{ {\bf k}, \nu }
)
\right],
\nonumber
\end{equation}
where $
\mathcal E_{ {\bf k}, \nu } 
=
\xi_{ {\bf k}, -} +(-1)^\nu \,
(\,
\xi_{{\bf k}, +}^2 + \vert \Delta_{\bf k} \vert^2
\,)^{1/2}
\nonumber
$ are the quasi-hole and quasi-particle energy spectrum for $\nu=1$ and $\nu =2$, respectively.
In addition, $\Delta_{\bf k} = \Delta_0 \Gamma_{\bf k}$ is the order parameter for superfluidity for pairing with zero center of mass momentum, 
$n_F ( E ) = 1/ (1+\exp E/T)$ is the Fermi distribution, and $
\xi_{{\bf k}, \pm}
= 
\left(
\xi_{{\bf k}, \uparrow} \pm 
\xi_{{\bf k}, \downarrow} 
\right)/2
= 
k^2/2m_{\pm} - \mu_\pm,
$ where $
m_{\pm} 
= 
2 m_{\uparrow} m_{\downarrow}
/
\left(
m_{\downarrow} \pm m_{\uparrow}
\right)
$ and $
\mu_{\pm}
=
\left(
\mu_{\uparrow}
\pm
\mu_{\downarrow}
\right)
/
2.
$ 
Notice that $m_{+}$ is twice the reduced mass of the $\uparrow$ and $\downarrow$ fermions, and that the equal mass case $( m_{\uparrow} = m_{\downarrow} )$ corresponds to $\vert m_{-} \vert \to \infty$.

The fluctuation term in the action leads to a correction to the thermodynamic potential, which can be written as 
\begin{equation}
\label{eqn:fermi-fermi-thermodynamic-potential}
\Omega_{G} 
= 
\Omega_0 
+ 
\Omega_{\rm fluct},
\end{equation}
where $\Omega_0 = T S_0$ and $
\Omega_{\rm fluct} 
= 
T
\sum_{q}
\ln {\rm det }
\left[
T {\bf F}^{-1} (q)
\right].
$

The saddle point condition $
\delta S_0/\delta \Delta_0^*
=
0
$ leads to the order parameter equation
\begin{equation}
\label{eqn:fermi-fermi-order-parameter}
\frac{1}{g}
= 
\sum_{{\bf k}}
\frac{ \vert \Gamma_{\bf k} \vert^2 } { 2 E_{ {\bf k} } }
\mathcal X_{ {\bf k} }^{-}.
\end{equation}
where $
E_{\bf k} =
(\,
\xi_{{\bf k}, +}^2 + \vert \Delta_{\bf k} \vert^2
\,)^{1/2}
$ and $
\mathcal X_{ {\bf k} }^{\pm}
=
n_F(\mathcal E_{{\bf k},1}) 
\pm 
n_F(\mathcal E_{{\bf k},2}).
$
As usual, we eliminate $g$ in favor of the scattering length $a_s$ via the relation 
\begin{equation}
\frac{1}{g} 
= 
- \frac{m_{+} V }{ 4\pi a_s }
+ 
\sum_{\bf k}
\frac{
\vert \Gamma_{\bf k} \vert^2
}{
\epsilon_{ {\bf k}, \uparrow }
+
\epsilon_{ {\bf k}, \downarrow }
}.
\nonumber
\end{equation}
The order parameter equation needs to be solved self-consistently with the number equations $
N_{\alpha} 
= 
- ( \partial \Omega / \partial \mu_{\alpha}  )_T,
$ which has two contributions
\begin{equation}
\label{eqn:fermi-fermi-number}
N_{\alpha} 
=
N_{0, \alpha}
+
N_{{\rm fluct}, \alpha}.
\end{equation}
Here, $
N_{0, \alpha}=
-( \partial \Omega_0 / \partial \mu_{\alpha} )_T
$ is the saddle point number equation given by
\begin{equation}
N_{0, \alpha} 
=
\sum_{\bf k}
\left(
\frac{1 - s_{\alpha}}{2}
-
\frac{\xi_{ {\bf k}, + } } {2 E_{{\bf k}} }  
\mathcal X_{ {\bf k}}^{-}
\right),
\end{equation}
where $
s_{\alpha} = \gamma_\alpha 
( 1 - 
\mathcal X_{ {\bf k}}^{+})
$ with $\gamma_\uparrow = 1$ and $\gamma_\downarrow = -1$.
And $
N_{{\rm fluct}, \alpha}
=
- 
(\partial \Omega_{\rm fluct} / \partial \mu_{\alpha} )\vert_T
$ is the fluctuation contribution to $N_{\alpha}$ given by 
\begin{equation}
N_{{\rm fluct}, \alpha }
=
-
T
\sum_{q}
\frac{
\partial }{
\partial \mu_{\alpha}} \,
\ln
{\rm det} \Big[ T {\bf F}^{-1} (q) \Big]
\end{equation}

In order to extract the isothermal compressibility $\kappa_T$ of the system or the bulk modulus $B = \kappa_T^{-1}$ shown in Eq.~(\ref{eqn:isothermal-compressibility}) and the pseudo-spin susceptibility $\chi_{zz}$ defined in Eq.~(\ref{eqn:pseudo-spin-susceptibility}), 
we need to obtain the pseudo-compressibility matrix $\{\tilde \kappa\}$ described in Eqs.~(\ref{eqn:pseudo-compressibility-1}) and~(\ref{eqn:fluctuation-dissipation}). 
Using the self-consistent solutions from the order parameter Eq.~(\ref{eqn:fermi-fermi-order-parameter}) and the number equation, Eq.~(\ref{eqn:fermi-fermi-number}), leads to a thermodynamic potential $\Omega$ which is a function of the order parameter $\Delta_0$, the chemical potentials $\mu_{\alpha}$, and temperature $T$. 
However, from a thermodynamic point of view $\Omega$ is ultimately dependent on $\mu_{\alpha}$ and $T$ only, such that $
\Omega 
\left[
\Delta_0 (\mu_{\alpha}, T),
\mu_{\alpha}(T), T
\right]
\to
\Omega 
\left[
\mu_{\alpha}(T), T
\right].
$
This implicit dependence of the order parameter $\Delta_0$ on the chemical potentials $\mu_\uparrow$, $\mu_\downarrow$, and temperature $T$ requires that derivatives of the thermodynamic potential $\Omega$ include both their explicit and implicit dependencies on chemical potentials, and temperature. 
This is extremely important for the calculation of the isothermal compressibility, spin susceptibility and pseudo-compressibility matrix to be discussed next.

\subsection{Pseudo-compressibility matrix without a trap}
\label{sec:fermi-fermi-pseudo-compressibility}
Because of the implicit dependence of $\Delta_0$ on $\mu_{\alpha}$, the direct calculation of the pseudo-compressibility requires
\begin{equation}
\frac{
{\tilde \kappa}_{\alpha \beta}
}{
T
}
=
\left( 
\frac{
\partial N_{\alpha}
}{
\partial \,\mu_{\beta} 
}
\right)_{T, {\rm e}}
+
\left(
\frac{
\partial \, N_{\alpha}
}{
\partial \vert \Delta_0 \vert^2
} 
\right)_{T, {\rm e}}
\cdot
\left(
\frac{
\partial \vert \Delta_0 \vert^2
}{
\partial \, \mu_\beta
} 
\right)_{T,{\rm i}},
\end{equation}
where the label ``e'' (``i'') means explicit (implicit) derivative.
The explicit expressions of the matrix elements at zero and finite temperature are given in Appendix~\ref{app-b}.
For any non-zero temperature $T$, the mechanical stability of the uniform superfluid and normal phases is guaranteed if all eigenvalues of $\tilde \kappa_{\alpha \beta}$ are positive, or, equivalently, if all eingenvalues of $\tilde \kappa_{ij}$ are positive. Since we are dealing with  $2 \times 2$ matrices, this is achieved when ${\rm Tr} \left[ \{\tilde \kappa\} \right] > 0$ and 
${\rm det} \left[ \{\tilde \kappa\} \right] > 0$. When the lowest eigenvalue of $\{\tilde \kappa\}$ reaches zero then the system becomes mechanically unstable. 
The existence of mechanical stability coincides with the condition that the thermodynamic potential is a minimum~\cite{iskin-06a} with respect to  variations of the order parameter $\vert \Delta_0 \vert^2$, this means that it satisfies simultaneously the extremum condition $(\partial \Omega/  \partial \vert \Delta_0 \vert^2) = 0$, and the positive curvature condition $(\partial^2 \Omega/\partial \vert \Delta_0 \vert^4) > 0$.
For zero temperature $(T = 0)$, the mechanical stability is guaranteed when both the eigenvalues of the matrix $\lim_{T \to 0} \left[ \{\tilde \kappa\}/T \right]$ are positive.

\begin{figure} [htb]
\centering
\includegraphics[width=1.0\linewidth]{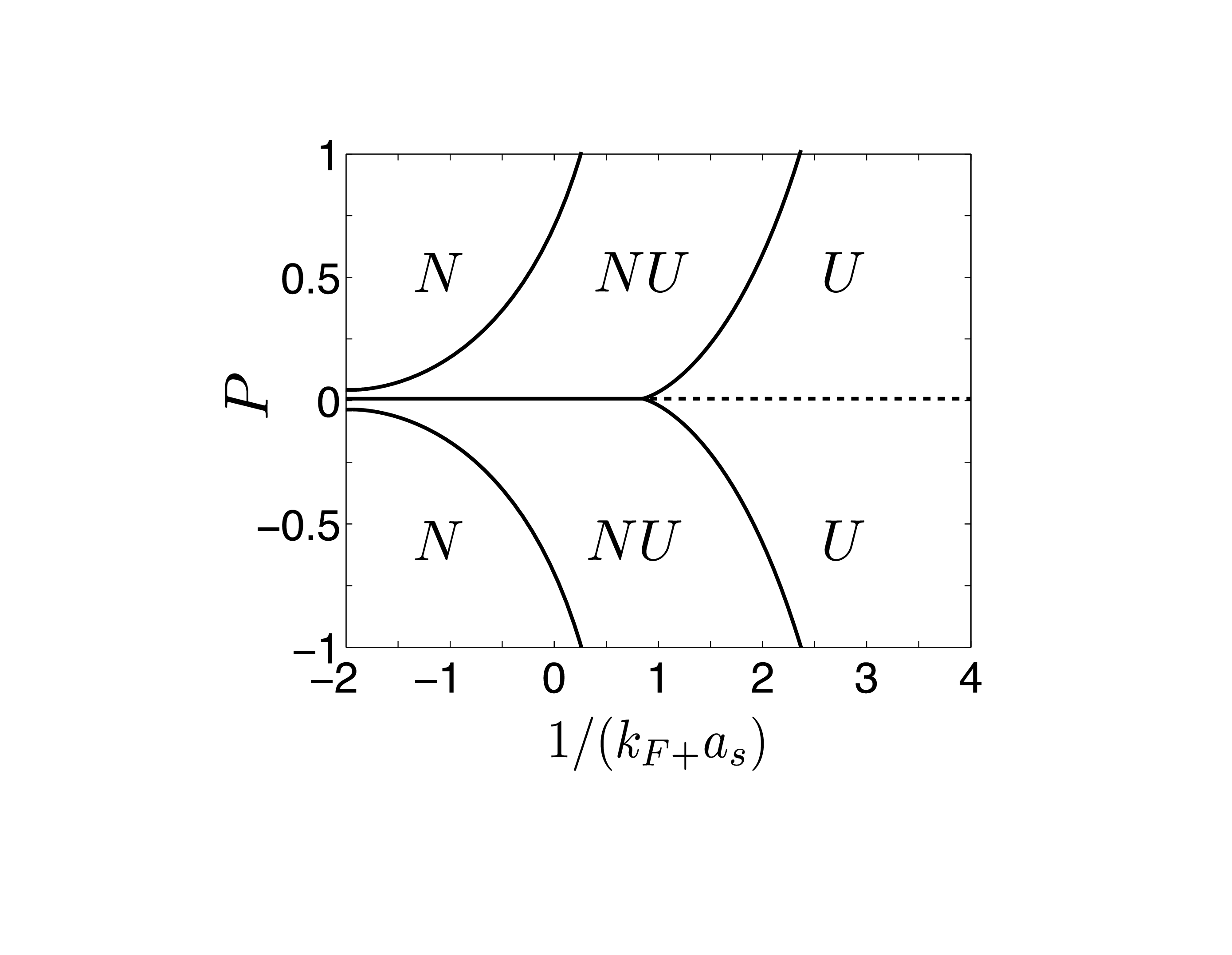}
\caption{ 
\label{fig:phase-diagram}
The zero temperature phase diagram of population imbalance $P = N_-/N_+$ versus scattering parameter $1/(k_{F+}a_{s})$ for equal mass fermions is  shown. 
For $P\ne 0$, the normal ($N$) state becomes superfluid (denoted by $NU$ and $U$) as the scattering parameter $1/(k_{F+}a_s)$ increases. 
$U$ represents the uniform superfluid phase, and $NU$ represents a non-uniform superfluid region, where the $U$ phase is unstable. 
For $P=0$, the $U$ phase is stable for all values of the scattering parameter. }
\end{figure}

Using the stability conditions described above in combination with the solutions of the order parameter Eq.~(\ref{eqn:fermi-fermi-order-parameter}) and number Eq.~(\ref{eqn:fermi-fermi-number}), we construct the zero temperature phase diagram for a mixture of fermions of equal masses, different hyperfine states and no trapping potential.
The zero temperature phase diagram of population imbalance $
P = ( N_{\uparrow} - N_{\downarrow} ) /  ( N_{\uparrow} +  N_{\downarrow} )
= 
N_{-} / N_{+}
$ versus scattering parameter $ 1/( k_{F +} a_s )$ is shown in Fig.~\ref{fig:phase-diagram}.
Here, $k_{F +}^3 = ( k_{F\uparrow}^3 + k_{F\downarrow}^3)/2$ is an effective Fermi momentum, where $k_{F\alpha}$ is the Fermi momentum of each species, and $\epsilon_{F+} = k_{F+}^2/(2m_{+})$ is an effective Fermi energy, which fixes the energy scale. 
The effective Fermi momentum $k_{F+}$ is convenient because it fixes the total number of particles $N_{+} = N_{\downarrow} + N_{\uparrow}$, where  the density for particle $\alpha$ is $(N_{\alpha} / V) = k_{F\alpha}^3/(6\pi^2)$.

\begin{figure} [htb]
\centering
\includegraphics[width=1.0\linewidth]{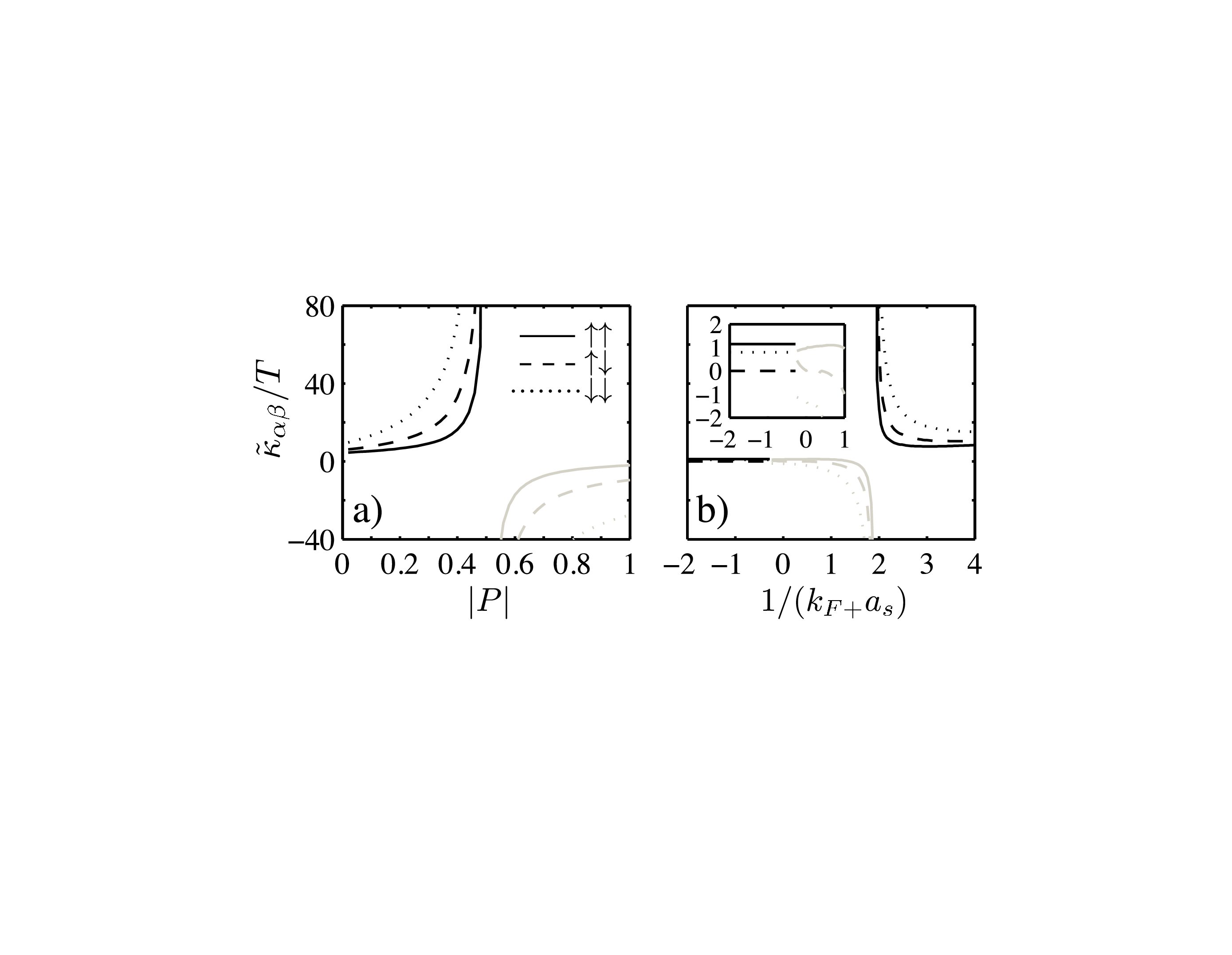}
\caption{ 
\label{fig:pseudo-compressibility-matrix}
The pseudo-compressibility matrix elements $\tilde \kappa_{\alpha \beta}/ T$ at $T=0$ are shown in a) as a function of population imbalance $P$ 
for interaction parameter $1/(k_{F+}a_s ) = 1.92$, and in b), as a function of $1/(k_{F+}a_s )$ for $P = 0.5$.
The matrix elements in the region where the uniform superfluid phase is unstable is shown in gray.
The inset in b) shows an expected discontinuity at the boundary between the normal and the non-uniform superfluid state. 
Notice that, in the normal state, $\tilde \kappa_{\uparrow \downarrow} = \tilde \kappa_{\downarrow \uparrow} = 0$. }
\end{figure}

In Fig.~\ref{fig:pseudo-compressibility-matrix}, we show the pseudo-compressibility matrix elements ${\tilde \kappa}_{\alpha \beta}$ in a) as a function of population imbalance $P$ for fixed $1/(k_{F_+} a_s) = 1.92$ and in b) as a function of $1/(k_{F_+} a_s)$ for fixed population imbalance $P = 0.5$. 
As seen in Fig.~\ref{fig:pseudo-compressibility-matrix}a, when population imbalance is changed in the BEC regime, e.g. $1/(k_{F_+} a_s ) = 1.92$,  
all $\lim_{T \to 0} {\tilde \kappa}_{\alpha \beta}/T$ diverge at the phase boundary between the uniform superfluid phase and the non-uniform phase as ${\tilde \kappa}_{\alpha \beta} \sim (P - P_c)^{-\nu}$ with $P_c = 0.5$ for $1/(k_{F+}a_s)=1.92$. 
The critical exponent associated with this divergence is $\nu = 1$ and can be extracted from the density fluctuations revealed in the matrix elements of ${\tilde \kappa}_{\alpha \beta}$.
In Fig.~\ref{fig:pseudo-compressibility-matrix}b, the matrix elements $\lim_{T \to 0}{\tilde \kappa}_{\alpha \beta}/T$ are shown as a function of  the interaction parameter $1/k_{F_+} a_s$ for fixed population imbalance $P = 0.5$.
As shown in the inset of Fig~\ref{fig:pseudo-compressibility-matrix}b, discontinuities are expected when the interaction parameter $1/( k_{F_+} a_s )$ is increased to cross the boundary between the normal phase and the region where a non-uniform superfluid phase is present.
The negative values for the matrix elements $\lim_{T \to 0}{\tilde \kappa}_{\alpha \beta}/T$ just indicate a region of non-uniform superfluidity,  that is, a region where uniform superfluidity is not mechanically stable. 
When the boundary between the non-uniform phase and the uniform superfluid phase is crossed then $\lim_{T \to 0}{\tilde \kappa}_{\alpha \beta}/T$ 
diverges as $(\lambda - \lambda_c)^{-1},$ where $
\lambda 
= 
1/(k_{F_+} a_s),
$ and $\lambda_c$ is the critical interaction parameter. For population imbalance $P  = 0.5$, the critical interaction parameter is $\lambda_c = 1.92$. 
This illustrates the point that using the generalized fluctuation-dissipation theorem described in Eqs.~(\ref{eqn:fluctuation-dissipation}) and~(\ref{eqn:compressibility-matrix}) allows for the extraction of critical exponents of density-density correlations accross phase boundaries.
Notice that in this approximation and within the normal state region, all $\lim_{T\to 0}\tilde \kappa_{\alpha\beta}/T$ are constant as a function  of interaction parameter $1/(k_{F+}a_s)$. 
Additional corrections to the present approximation are necessary to capture fully interaction effects in the normal state.

\subsection{Isothermal Compressibility and Spin Susceptibility}
\label{sec:isothermal-compressibility-spin-susceptibility}

\begin{figure} [htb]
\centering
\includegraphics[width=1.0\linewidth]{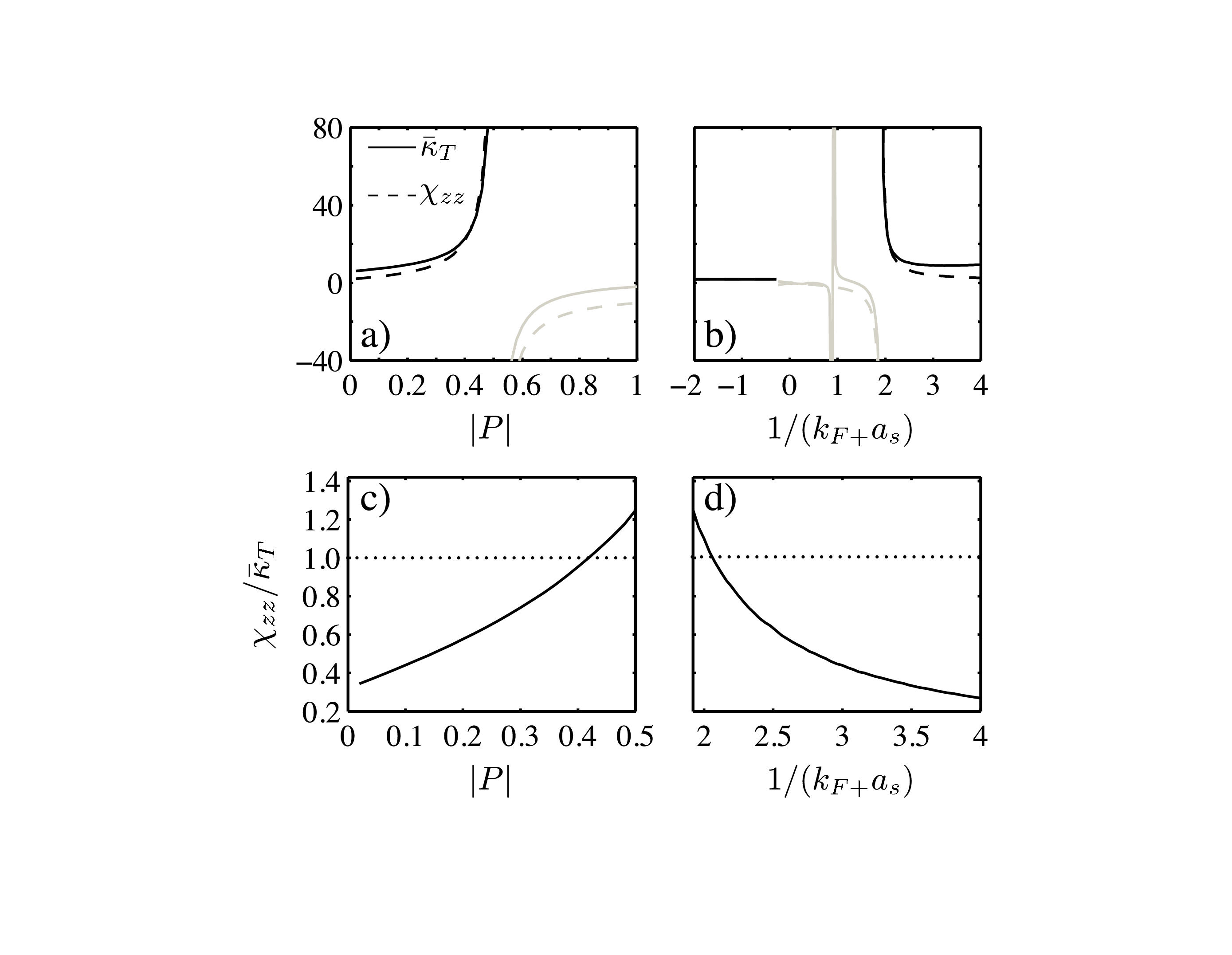}
\caption{ 
\label{fig:isothermal-compressibility}
The isothermal compressibility per unit volume $V$, ${\bar \kappa}_T = \kappa_T/V$ as well as the pseudo-spin susceptibility $\chi_{zz} = \tilde \kappa_{--}/T$ in the limit of $T=0$ are shown in a) as a function of $P$ for $1/(k_{F+}a_s)=1.92$, and in b) as a function of $1/(k_{F+}a_s)$ for $P = 0.5$.
The unit on the vertical axis is chosen as $3N_+/4\epsilon_{F+}$. 
The gray curves in a) and b) illustrate the instability region of the uniform superfluid phase. 
In c) and d), the ratio $\chi_{zz} / \kappa_T$ is shown as a function of $P$ and $1/(k_{F+}a_s)$ in a stable uniform superfluid state, respectively. }
\end{figure}

The isothermal compressibility per unit volume ${\bar \kappa}_T = \kappa_T/V$ and the pseudo-spin susceptibility $\chi_{zz}$ are shown in Fig.~\ref{fig:isothermal-compressibility}a as a function of population imbalance $P$ for $1/(k_{F_+} a_s) = 1.92$ as in Fig.~\ref{fig:pseudo-compressibility-matrix}.
Notice that as $P$ increases, both $\kappa_T$ and $\chi_{zz}$ diverge as $(P - P_c )^{-\nu}$ at the critical population imbalance $P_c = 0.5$, and become negative for $|P| > P_c$ signaling a quantum phase transition from uniform superfluidity with coexistence of excess unbound fermions and paired fermions in the same spatial region into a phase separated regime where excess unbound fermions and paired fermions tend to avoid being in the same region of space. 
The critical exponent in the region where the isothermal compressibility and the pseudo-spin susceptibility are positive has the saddle point (mean field) value of $\nu = 1$. 
A renormalization group calculation would be necessary to obtain corrections to this critical exponent, but we postpone it to a later date.

In Fig.~\ref{fig:isothermal-compressibility}b, we show ${\bar \kappa}_T$ and $\chi_{zz}$ as a function of $1/(k_{F_+} a_s)$ for fixed population imbalance $P = 0.5$. 
In the uniform superfluid phase, where $1/(k_{F+}a_s) > \lambda_c$, the same divergent behavior occurs as $\kappa_T \sim (\lambda - \lambda_c)^{-1}$ and $\chi_{zz} \sim (\lambda - \lambda_c)^{-1}$, where $\lambda = 1/(k_{F_+} a_s)$ and $\lambda_c$ is the critical interaction parameter ($\lambda_c = 1.92$ for $P =0.5$).  
The critical exponent $\nu = 1$ reflects the saddle point approximation used to calculate it. Renormalization group calculations are necessary to obtain corrections to $\nu = 1$.

In Fig.~\ref{fig:isothermal-compressibility}c, we show the ratio of the pseudo-spin susceptibility $\chi_{zz}$ to the isothermal compressibility 
per unit volume ${\bar \kappa}_T$
\begin{equation}
\frac{\chi_{zz}}{{\bar \kappa}_T}
=
N_{+}^2
\left[
\left( \frac{\tilde \kappa_{--}}{\tilde \kappa_{++}} \right) + 
2 P \left(  \frac{\tilde \kappa_{--}}{\tilde \kappa_{+-}} \right) + 
P^2 \right]
\end{equation}
as a function of $|P|$ with a fixed value of $1/(k_{F+}a_s) = 1.92$, and Fig.~\ref{fig:isothermal-compressibility}d shows the ratio as a function   
$1/(k_{F+}a_s)$ for $P = 0.5$ in the uniform superfluid phase.
In Figs.~\ref{fig:isothermal-compressibility}c and~\ref{fig:isothermal-compressibility}d notice that the ratio $\chi_{zz}/{\bar \kappa}_T$ approaches a finite value, as indicated by the dotted line, reflecting the same power law divergence of ${\bar \kappa}_T$ and $\chi_{zz}$ at the phase boundary where the uniform $(U)$ superfluid phase becomes unstable.

\begin{figure} [htb]
\centering
\includegraphics[width=1.0\linewidth]{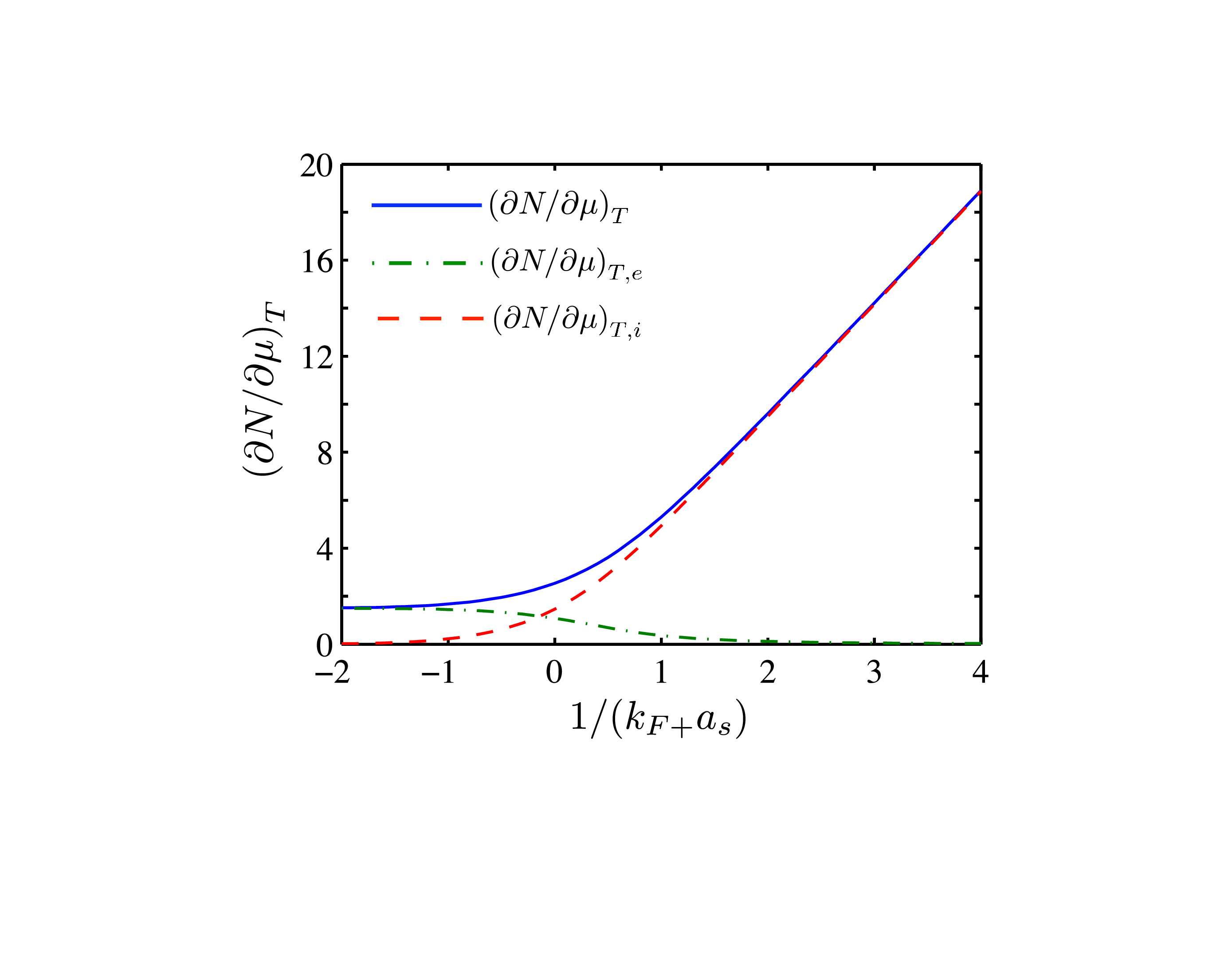}
\caption{ 
\label{fig:isothermal-compressibility-balanced}
(color online) $(\partial N / \partial \mu)_T$ as a function of $1/(k_{F+}a_s)$ for $T=0$
is plotted by the blue solid line
for the population balanced case $P = 0$.
The unit on the vertical axis is chosen as $N/\epsilon_F$. 
The green (dot-dashed) line represents the contribution 
from the explicit derivative $(\partial N / \partial \mu)_{T,{\rm e}}$,
while the red (dashed) line corresponds to the contribution from 
the implicit dependence of
$|\Delta_0|$ on $\mu$, that is, from 
$(\partial N / \partial \mu)_{T, {\rm i}} = 
(\partial N / \partial |\Delta_0|^2 )_{T,{\rm e}} \cdot
(\partial |\Delta_0|^2 / \partial \mu )_{T,{\rm i}}$.
}
\end{figure}

For completeness, in Fig.~\ref{fig:isothermal-compressibility-balanced}, we show plots of the zero-temperature isothermal compressibility for the  population balanced case corresponding to $P = 0$. 
In particular, we show separately the contributions coming from both the explicit and implicit dependencies on the chemical potential $\mu$ in the expression for the isothermal compressibility
\begin{equation}
\kappa_T
= 
\frac{V}{T} 
\left(
\frac{\tilde \kappa}{N^2}
\right)
=
\frac{V}{N^2}
\left(
\frac{\partial  N}
{\partial \,\mu}
\right)_T,
\end{equation}
where $N = \langle \hat N \rangle$ is the total average number of particles, and $\tilde \kappa = T (\partial  N / \partial \mu )_T$ is the $1 \times 1$ compressibility matrix for the balanced case, that is, the scalar 
\begin{equation}
\label{eqn:compressibility-balanced-explicit-implicit}
\frac{
\tilde \kappa 
}{ T}
=
\left(
\frac{\partial N} {\partial \, \mu } 
\right)_{T, {\rm e}}
+
\left(
\frac{\partial N} {\partial \, \mu } 
\right)_{T, {\rm i}},
\end{equation}
where the implicit derivative is 
\begin{equation}
\left(
\frac{\partial N} {\partial \, \mu } 
\right)_{T, {\rm i}}
=
\left(
\frac{\partial \, N} {\partial \vert \Delta_0 \vert^2} 
\right)_{T, {\rm e}}
\cdot
\left(
\frac{\partial \vert \Delta_0 \vert^2} {\partial \, \mu} 
\right)_{T,{\rm i}}.
\end{equation}

As seen in Fig.~\ref{fig:isothermal-compressibility-balanced}, the second term in Eq.~(\ref{eqn:compressibility-balanced-explicit-implicit}), which contains the implicit dependence on the chemical potential $\mu$, is important both in the intermediate region $-1 < 1/(k_{F}a_s) < 1$ around unitarity and in the BEC regime. Analytical limits in the BCS and BEC regimes are also shown and confirm the importance of the second term. 
The derivations are given in Appendix~\ref{app-c}.
The zero-temperature behavior of $\tilde \kappa /T = (\partial N/\partial \mu )_T$ in the BCS limit is given by
\begin{equation}
\left(
\frac{\partial N}
{\partial \mu} 
\right)_T
\simeq
\frac{3N}{2\epsilon_F}
\left(
1
+
\gamma \, e^{-\frac{\pi}{\vert k_F a_s \vert}}
\right),
\end{equation}
having a positive correction with an upward curvature as the interaction parameter increases: $\gamma = [ 7\sqrt{2} - 4 + 3 \sinh^{-1}(1) ]\,8/e^4 \simeq 1.2519$.
The corresponding result in the BEC limit is given by 
\begin{equation}
\frac{\partial N}
{\partial \mu}
\simeq
\frac{3N}{2\epsilon_F}\,
\frac{\pi}
{k_F a_s},
\end{equation}
which agrees with the result of the compressibility of a weakly interacting gas of bosons with mass $m_B = 2m$, and scattering parameter $a_B = 2 a_s$~\cite{huang-book}.
Thus, it is only when the implicit dependence of the order parameter $\Delta_0$ on the chemical potential $\mu$ that the qualitatively correct result in the BEC limit is recovered. This can be generally understood by the argument that increasingly attractive interactions tend to make the  Fermi system less degenerate and thus more compressible.
This tendency is better illustrated in the BEC regime where molecular bosons emerge as new degrees of freedom and the Pauli pressure is weakened with increasing interaction parameter thus leading to a more compressible Fermi superfluid.

Before we conclude this section it is worth noting that the zero-temperature spin susceptibility $\chi_{zz}$ is exactly zero in the case of balanced populations $(P = 0)$, since for a uniform $s$-wave superfluid at $T = 0$ all the fermions are paired.

Having discussed the isothermal compressibility and spin-susceptibility for $s$-wave Fermi superfluids without a trap, now we turn our attention 
to the case where the trapping potential is non-zero.

\section{Fermi-Fermi mixtures in a trap}
\label{sec:fermi-fermi-mixtures-trap}
In the case of a non-zero trapping potential $V_{\alpha} ({\bf r}),$ we make use of the local density approximation (LDA), and obtain the local thermodynamic potential 
\begin{equation}
\label{eqn:local-thermodynamic-potential}
\Omega ({\bf r}) 
= 
\Omega 
\left[ 
\mu_{\alpha} ({\bf r})
\right]
\end{equation}
from the thermodynamic potential in the absence of a trap, defined in Eq.~(\ref{eqn:fermi-fermi-thermodynamic-potential}), via the substitution 
\begin{equation}
\label{eqn:local-chemical-potential}
\mu_{\alpha} 
\to 
\mu_{\alpha} ({\bf r}) 
= 
\mu_{\alpha} - V_{\alpha} ({\bf r}).
\end{equation}
We note, however, that the generalized fluctuation dissipation theorem described in Sec.~{\ref{sec:fermi-fermi-mixtures-trap}} can be used to extract the local compressibility and the local spin-susceptibility directly from experimental results of density-density fluctuations, without invoking the local density approximation (LDA).

\subsection{Density profiles and order parameter}
\label{sec:fermi-fermi-density-order-parameter}
The use of LDA implies that both the order parameter $\Delta_0$ and the number of particles $N_{\alpha}$ defined in Eqs.~(\ref{eqn:fermi-fermi-order-parameter}) and~(\ref{eqn:fermi-fermi-number}), respectively, become functions of position ${\bf r}$ via the position dependent chemical 
potentials $\mu_{\alpha} ({\bf r})$. 
As a result, we have $
\Delta_0 ({\bf r}) 
= 
\Delta_0 
\left[
\mu_{\uparrow} ( {\bf r} ),\mu_{\downarrow} ( {\bf r} )
\right]
$ and $
n_{\alpha} ({\bf r})
=
n_{\alpha}
\left[
\mu_{\uparrow} ( {\bf r} ),\mu_{\downarrow} ( {\bf r} )
\right],
$ where $n_{\alpha} ({\bf r}) = N_{\alpha} ({\bf r})/V$ is the local density for fermions of type $\alpha$. 
To simplify our discussion, we consider isotropic harmonic trapping potentials 
\begin{equation}
V_{\alpha} ({\bf r}) 
= 
\frac{\gamma_{\alpha}}{2} 
r^2,
\end{equation}
where $\gamma_{\alpha} = m_{\alpha} \omega_{\alpha}^2$ with $\omega_{\alpha}$ being the trapping frequency of fermion of type $\alpha$, and $r = \vert {\bf r} \vert$ is the magnitude of the position vector ${\bf r}$.

\begin{figure} [htb]
\centering
\includegraphics[width=1.0\linewidth]{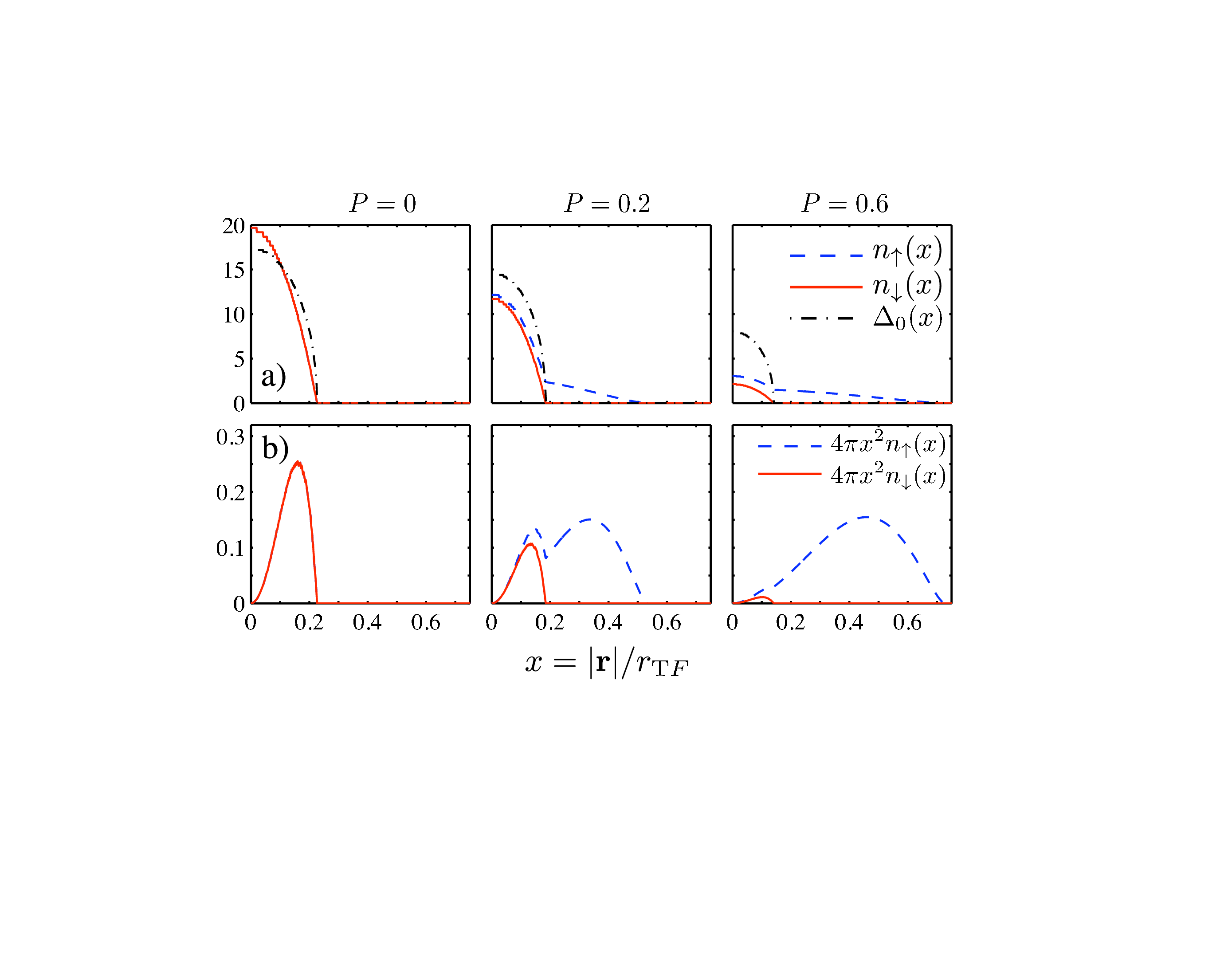}
\caption{ 
\label{fig:spatial-dependence-density}
(color online) 
a) The particle density profiles $n_{\alpha}(x)$ and b) the Jacobian-weighted $4 \pi x^2 n_{\alpha}(x)$ normalized by $N_+ = (2m_+\epsilon_{F+})^{3/2} / (3 \pi^2)$, are plotted as a function of dimensionless position $x = \vert \mathbf{r} \vert /r_{\rm{TF}}$ for fixed interaction parameter  $1/(k_{F+}a_s) = 3.0$ at the population imbalances $P=0,$ $P=0.2$, and $P=0.6$, respectively. 
The dimensionless order parameter $\Delta_0 (x)$ normalized by $\epsilon_{F+}$ is represented by the black-dot-dashed line, while the densities $n_{\uparrow} (x)$ and $n_{\downarrow} (x)$ are described by the blue-dashed and red-solid lines, respectively. 
Notice that for $P = 0$, the densities $n_{\uparrow} (x)$ and $n_{\downarrow} (x)$ coincide exactly. }
\end{figure}

Following the general procedure of Sec.~\ref{sec:compressibility-trap}, it is simple to show that within LDA the local density is given by 
\begin{equation}
n_{\alpha} ({\bf r})
=
-
\frac{ \partial\Omega [ \mu_\uparrow ({\bf r}), \mu_\downarrow ({\bf r}) ]}
{\partial \, \mu_\alpha ({\bf r})},
\end{equation}
where the thermodynamic potential $\Omega$ is defined in Eq.~(\ref{eqn:fermi-fermi-thermodynamic-potential}) with the replacement $\mu_{\alpha} \to \mu_{\alpha} ({\bf r})$, where $\mu_{\alpha} ({\bf r})$ is the local chemical potential defined in Eq.~(\ref{eqn:local-chemical-potential}).

For the equal mass case, we show in Fig.~\ref{fig:spatial-dependence-density}a the spatial dependence of the particle density profiles $n_{\alpha} (x)$ and the order parameter $\Delta_0 (x)$ as a function of dimensionless position $x = |{\bf r}|/r_{TF}$, where $r_{\rm TF}$ is the Thomas-Fermi radius defined through the condition $\epsilon_{F+} = \gamma_{+} r_{\rm TF}^2/2$, where $\gamma_{+} = \gamma_{\uparrow} + \gamma_{\downarrow}$.
These spatial profiles show that superfluidity coexists with excess unpaired fermions, e.g. spin-up fermions $n_\uparrow (x)$, but the majority of excess unpaired fermions are pushed away from the center of the trap.  This effect is better seen in Fig.~\ref{fig:spatial-dependence-density}b, where the Jacobian-weighted particle numbers $4\pi x^2 \, n_{\alpha} ({ x})$ are shown for interaction parameter $1/(k_{F+} a_s) = 3.0$ at population imbalances of $P = 0$, $P = 0.2$, and $P= 0.6$.

Given our choice of majority spin-up fermions, the density of paired fermions $n_{\text{pair}} ({\bf r})$ is controlled by the minority spin-down  fermions, becoming $n_{\text{pair}}({\bf r}) = 2 n_{\downarrow} ({\bf r})$, and the density of excess fermions $n_\text{e} ({\bf r})$ is simply 
the difference between the densities of the spin-up and spin-down fermions such that $n_\text{e} ({\bf r}) = n_{\uparrow} - n_{\downarrow} ({\bf r})$. 
Notice that $n_{\text{pair}} ({\bf r})$ vanishes beyond the critical radius ${\bf r}_c$ beyond which the order parameter for superfluidity $\Delta_0 ({\bf r})$ is zero.
Furthermore, as population imbalance $P$ increases the density of pairs $n_{\text{pair}} ({\bf r})$ also decreases concomitantly with a shrinkage  of the superfluid region determined by $r_c = \vert {\bf r}_c \vert$. In contrast, as $P$ increases, the overall density of excess fermions $n_\text{e} ({\bf r})$ increases.

\begin{figure} [htb]
\centering
\includegraphics[width=1.0\linewidth]{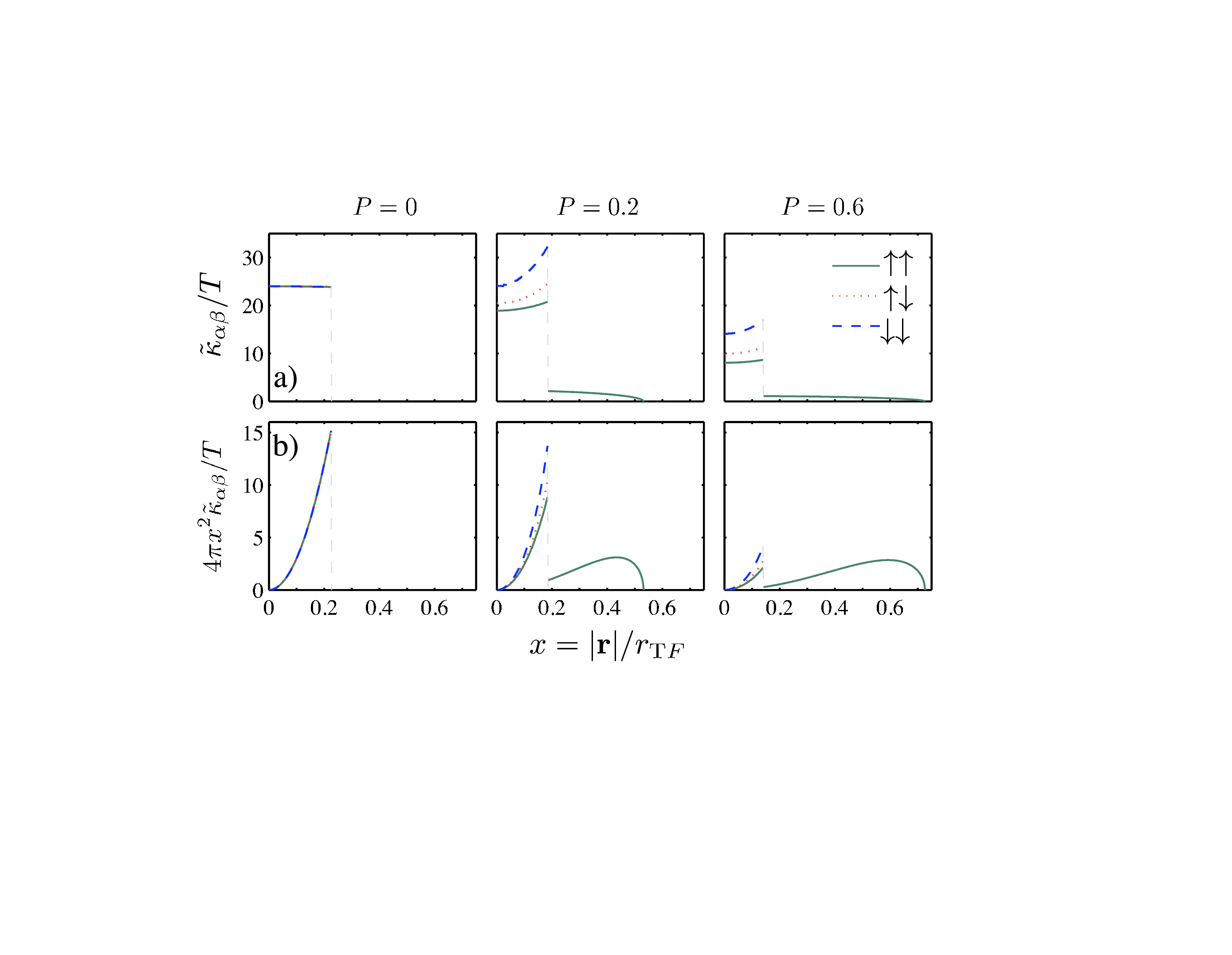}
\caption{
\label{fig:spatial-dependence-compressibility-matrix}
(color online) The profile of pseudo-compressibility matrix elements $\tilde \kappa_{\alpha\beta}(x)/T$ are shown in a) as a function of dimensionless position $x$ for fixed interaction parameter $1/(k_{F+}a_s) = 3.0$ at population imbalances $P = 0$, $P=0.2$, and $P=0.6$.
And the corresponding Jacobian-weighted profiles $4 \pi x^2 \kappa_{\alpha\beta}(x)$ are shown in b). 
The unit on the vertical axis is chosen as $3 N_+ / 4 \epsilon_{F+}$.
The solid-green line denotes the $\uparrow\uparrow$ element of both pseudo-compressibility and Jacobian-weighted matrix.
The dotted-red line describes ${\uparrow\downarrow}$ element, and the dashed-blue line describes ${\downarrow\downarrow}$ element. }
\end{figure}

\subsection{Compressibility and spin susceptibility}
\label{sec:fermi-fermi-compressibility-spin-susceptibility}
Following the steps outlined in Sec.~\ref{sec:compressibility-trap}, it is straightforward to show that the matrix elements of the pseudo-compressibility matrix can be obtained from the relation 
\begin{equation}
\tilde \kappa_{\alpha \beta} ({\bf r}) 
=
- T
\frac{ \partial^2 \Omega [\mu_{\uparrow} ({\bf r}),\mu_{\downarrow} ({\bf r})] }
{\partial \mu_{\alpha} ({\bf r}) \partial \mu_{\beta} ({\bf r})},
\end{equation}
where $\Omega [\mu_{\uparrow} ({\bf r}), \mu_{\downarrow} ({\bf r})]$ is the thermodynamic potential within LDA.
In Fig.~\ref{fig:spatial-dependence-compressibility-matrix}a, the spatial dependence of the pseudo-compressibility matrix elements $\tilde \kappa_{\alpha \beta} ({x}) / T$ are shown for the equal mass case as a function of dimensionless position ${x}$ for the same parameters used 
in Fig.~\ref{fig:spatial-dependence-density}. 
Notice that for $P = 0$ all the matrix elements of $\tilde \kappa_{\alpha \beta} ({x}) / T$ in the superfluid phase, $x < x_c$, coincide and vanish at $x > x_c$. 
For $P \ne 0$, however, the matrix elements are split at $x<x_c$, reflecting the coexistence of the superfluidity and excess unpaired fermions, and vanish at $x>x_c$ except for $\tilde \kappa_{\uparrow\uparrow}(x)/T$ due to the unpaired spin-up fermions. 
In Fig.~\ref{fig:spatial-dependence-compressibility-matrix}b, the Jacobian-weighted pseudo-compressibility matrix elements $4\pi x^2 \tilde \kappa_{\alpha \beta} ({x}) / T$ are plotted, showing that these matrix elements decrease in the superfluid region $(x < x_c)$ as $P$ increases, since excess fermions are transferred to the non-superfluid regions $(x > x_c)$.
Notice, however, that $\tilde \kappa_{\uparrow\uparrow}(x)/T$ in the normal phase beyond $x_c$ increases with $P$ for the same reason.
Within LDA, each matrix element exhibit a significant discontinuous drop when crossing the critical radius ${x}_c$.

\begin{figure} [htb]
\centering
\includegraphics[width=1.0\linewidth]{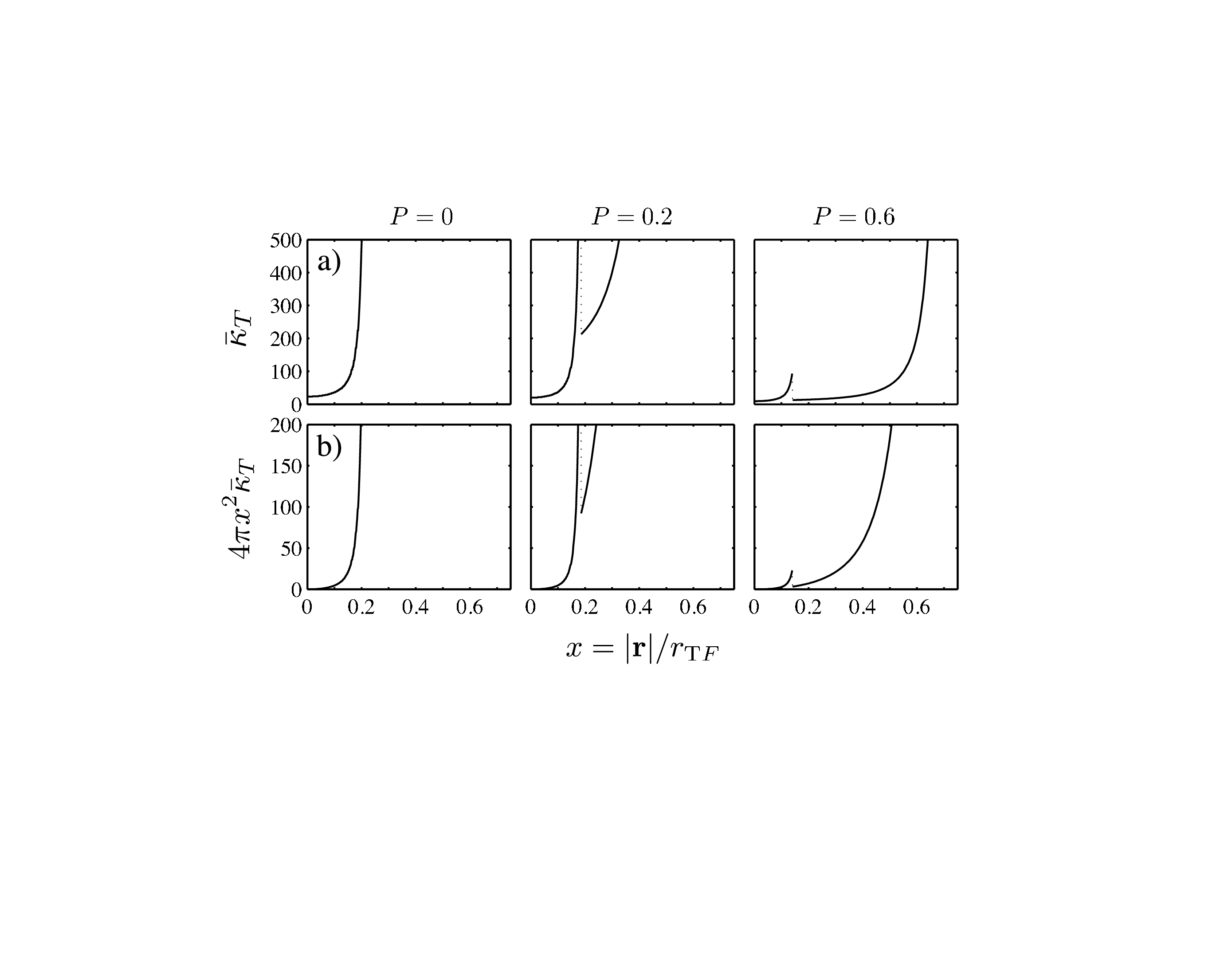}
\caption{ 
\label{fig:spatial-dependence-isothermal-compressibility}
a) The zero-temperature isothermal compressibility profile per unit volome $\bar \kappa_T (x)$ and b) the Jacobian-weighted 
profile $4 \pi x^2 \bar \kappa_T (x) $ in unit of $3N_+/ 4\epsilon_{F+}$ are shown as a function of dimensionless position $x$ for fixed interaction parameter $1/(k_{F+}a_s) = 3.0$ at population imbalances $P = 0$, $P=0.2,$ and $P=0.6$. }
\end{figure}

\begin{figure} [htb]
\centering
\includegraphics[width=1.0\linewidth]{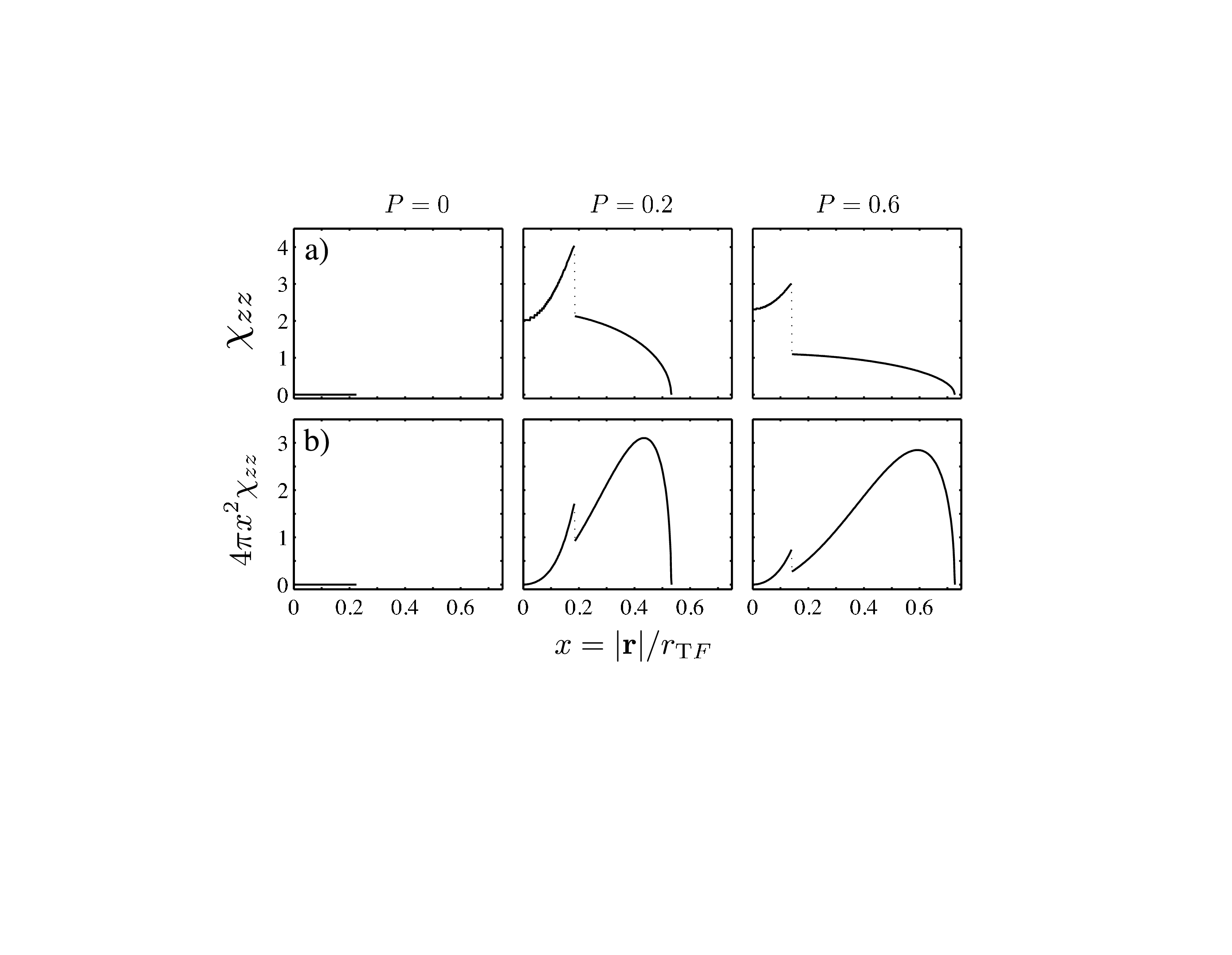}
\caption{ 
\label{fig:spatial-dependence-spin-susceptibility}
a) The zero-temperature pseudo-spin susceptibility profile $\chi_{zz} (x) = \tilde \kappa_{--} (x) / T$ and b) the Jacobian-weighted 
profiles $4 \pi x^2 \chi_{zz} (x)$ in unit of $3N_+/ 4\epsilon_{F+}$ is shown as a function of dimensionless position $x$ for fixed interaction parameter $1/(k_{F+}a_s) = 3.0$ at population imbalances $P = 0$, $P=0.2,$ and $P=0.6$. }
\end{figure}

In Figs.~\ref{fig:spatial-dependence-isothermal-compressibility} and \ref{fig:spatial-dependence-spin-susceptibility}, the spatial dependence of the isothermal compressibility $\kappa_T (x)$ and that of the spin-susceptibility $\chi_{zz} (x)$ are shown, respectively, for the same parameters as Fig.~\ref{fig:spatial-dependence-density}.
As seen in Fig.~\ref{fig:spatial-dependence-isothermal-compressibility}, $\kappa_T (x)$ exhibits a discontinuity at the position $x = x_c$ where the order parameter $\Delta_0 (x)$ and the minority density $n_{\downarrow} (x)$ vanish. 
Since, within LDA, the isothermal compressibility is 
\begin{equation}
\label{eqn:isothermal-compressibility-lda}
\frac{1}{\kappa_T (x)} 
=
\frac{T}{V}
\sum_{\alpha\beta} 
\left(
\frac{ n_{\alpha} (x) n_\beta (x) } 
{\tilde\kappa_{\alpha\beta}(x)}
\right), 
\end{equation}
the discontinuity becomes evident as $n_{\downarrow}(x) \to  0$, when $x \to x_c$.
For finite population imbalance $P$, $\kappa_T(x)$ increases monotonically with $x$ in the superfluid region $x<x_c$, and exhibits a discontinuity at $x_c$.
For the population balanced case $P = 0$, the isothermal compressibility $\kappa_T (x)$ also increases monotonically with $x$ and exhibits a discontinuity at the boundary between the superfluid region and vacuum (where there are no fermions). 
These discontinuities at $x = x_c$ are also clearly seen in Fig.~\ref{fig:spatial-dependence-isothermal-compressibility}b, where the Jacobian-weighted isothermal compressibility profiles $4 \pi x^2 \kappa_T(x)$ are shown.

As seen in Fig.~\ref{fig:spatial-dependence-spin-susceptibility}, for non-zero $P$, the pseudo-spin susceptibility $\chi_{zz}(x)$ also 
exhibits a discontinuity at the position $x_c$, where the order parameter $\Delta (x)$ and the minority density $n_{\downarrow} (x)$ vanish.
This discontinuity signals the phase boundary between the spatial region where superfluidity coexists with excess fermions, and the spatial region where only excess fermions exist.
In Fig.~\ref{fig:spatial-dependence-spin-susceptibility}b the Jacobian-weighted pseudo-spin susceptibility profiles $4 \pi x^2 \chi_{zz}(x)$ are shown.
Notice that the overall response $\chi_{zz} = (r_{TF}^3/V) \int dx 4 \pi x^2 \chi_{zz}(x)$ increases monotonically with $P$, as more unbound excess fermions are available.
In the balanced case, where $P = 0$, we have trivially $\chi_{zz} (x) = 0$, since for $x <x_c$ all spin-$\uparrow$ and spin-$\downarrow$ fermions 
are paired into a singlet state, and for $x  > x_c $ there are no fermions.

Now that we have completed our discussion of the spatial dependence of the particle density, order parameter, compressibility and spin susceptibility for balanced and imbalanced Fermi-Fermi mixtures, we summarize our main results and state our conclusions.

\section{Summary and Conclusions}
\label{sec:summary-conclusions}
We derived general relations connecting the isothermal compressibility and the spin susceptibility to fluctuations in the particle numbers for mixtures of equal mass fermions with population imbalance and for unequal mass fermions. This derivation produced a generalized fluctuation-dissipation theorem for Fermi-Fermi mixtures, that can also be applied to any other binary mixtures such as Bose-Fermi or Bose-Bose mixtures, and  that is also valid beyond the local density approximation.
The generalized fluctuation-dissipation theorem was used to connect theoretical calculations of the isothermal compressibility and spin susceptibility to experimentally accessible density and density fluctuation profiles of balanced and imbalanced mixtures of equal mass fermions. 
We described the isothermal compressibility and spin susceptibility for translationally invariant continuum systems, as non-translationally invariant continuum systems in the presence of a trapping potential. 
Using the local density approximation, we obtained expressions relating the local compressibility and local spin susceptibility to local fluctuations in particle numbers, and described their spatial profiles as a function of population imbalance. 
Lastly, we argued that discontinuities in these thermodynamic quantities can be used to identify the phase boundaries between two qualitative different phases such as the superfluid and normal states.

\begin{acknowledgements}
We would like to thank the Army Research Office (Contract No. W911NF-09-1-0220) for support.
\end{acknowledgements}

\appendix
\section{INVERSE FLUCTUATION PROPAGATOR}
\label{app-a}
In this appendix, we present the elements of the inverse fluctuation propagator matrix ${\bf F}^{-1} ({\bf q},i\omega_n)$, where $\omega_n = 2 \pi n T$ is the bosonic Matsubara frequency at temperature $T$.
The diagonal matrix element reads
\begin{equation}
{\bf F}^{-1}_{11}
= 
\frac{1}{g} +
\sum_{\bf k} |\Gamma_{\bf k}|^2 U_{{\bf q}/2,{\bf k}}
\left( 
T_{{\bf q}/2,{\bf k}}^{\rm intra} + T_{{\bf q}/2,{\bf k}}^{\rm inter} 
\right),
\end{equation}
and the off-diagonal matrix element
\begin{equation}
{\bf F}^{-1}_{12}
=
\sum_{\bf k} |\Gamma_{\bf k}|^2 
V_{{\bf q}/2, {\bf k}}\,
\left( 
T_{{\bf q}/2,{\bf k}}^{\rm intra} - T_{{\bf q}/2,{\bf k}}^{\rm inter} 
\right),
\end{equation}
where $U_{{\bf q},{\bf k}}$ and $V_{{\bf q},{\bf k}}$ are the coherence factors 
\begin{equation}
U_{{\bf q},{\bf k}} 
= 
\frac{1}{4}
\left( 1 + \frac{\xi_{{\bf q} + {\bf k},+} }{ E_{{\bf q} + {\bf k}} }\right) 
\left( 1 - \frac{\xi_{{\bf q} - {\bf k},+} }{ E_{{\bf q} - {\bf k}} }\right),
\end{equation}
\begin{equation}
V_{{\bf q}, {\bf k}}
= 
\frac{|\Delta_{{\bf q} + {\bf k}}||\Delta_{{\bf q} - {\bf k}}| }{
4 E_{{\bf q} + {\bf k}}E_{{\bf q} - {\bf k}}},
\end{equation}
and the intra-band and inter-band transfers
\begin{equation}
T_{{\bf q},{\bf k}}^{\rm intra}
=
A^{11}_{{\bf q}/2,{\bf k}}
+
A^{22}_{{\bf q}/2,{\bf k}},
\end{equation}
\begin{equation}
T_{{\bf q},{\bf k}}^{\rm inter} 
=
A^{12}_{{\bf q}/2,{\bf k}}
+
A^{21}_{{\bf q}/2,{\bf k}},
\end{equation}
with
\begin{equation}
A^{\nu \nu^\prime}_{{\bf q},{\bf k}}
=
\frac{ 
n_F(\mathcal E_{{\bf q} - {\bf k},\nu}) 
- 
n_F(\mathcal E_{{\bf q} + {\bf k},\nu^\prime})}
{ i\omega_n
+ 
\mathcal E_{{\bf q} - {\bf k},\nu} 
- 
\mathcal E_{{\bf q} + {\bf k},\nu^\prime}  }.
\end{equation}

\section{
Compressibility Matrix Elements $\kappa_{\alpha\beta}$ 
}
\label{app-b}
In this appendix, we present the explicit expression of the pseudo-compressibility matrix element $\kappa_{\alpha\beta}$, which is a partial derivative of $N_\alpha$ with respect to $\mu_\beta$ with fixed $T$.
Since $|\Delta_0|$ depends on $\mu_\beta$ implicitly, we need to add an implicit partial derivative of $|\Delta_0|$ with respect to $\mu_\beta$. 
\begin{equation}\label{b1}
\frac{\tilde\kappa_{\alpha \beta} }{ T } 
= 
\left( \frac{\partial N_\alpha}{\partial \mu_{\beta}} \right) + 
\left( \frac{\partial N_\alpha}{\partial |\Delta_0|^2} \right)
\left( \frac{\partial |\Delta_0|^2}{\partial \mu_{\beta}} \right)
\end{equation}
But, since the uniform superfluid phase satisfy the order parameter equation $(\partial \Omega / \partial |\Delta_0|^2 ) = 0$ and $N_\beta = -(\partial \Omega / \partial \mu_\beta )$ simultaneously.
Taking a partial derivative of order parameter equation with respect to $\mu_\beta$, we have
\begin{equation}\nonumber
\frac{\partial }{\partial \mu_{\beta}} 
\left( \frac{\partial \Omega}{\partial |\Delta_0|^2} \right) + 
\left( \frac{\partial^2 \Omega}{\partial |\Delta_0|^4} \right)
\left( \frac{\partial |\Delta_0|^2}{\partial \mu_{\beta}} \right)= 0.
\end{equation}
The first term can be rewritten as $-(\partial N_\beta / \partial |\Delta_0|^2 )$, leading to 
\begin{equation}\nonumber
\left( \frac{\partial |\Delta_0|^2}{\partial \mu_{\beta}} \right) = 
\left( \frac{\partial N_{\beta}}{\partial |\Delta_0|^2} \right)
\left( \frac{\partial^2 \Omega}{\partial |\Delta_0|^4} \right)^{-1}.
\end{equation}
Thus, the implicit partial derivative of $|\Delta_0|$ in Eq.~(\ref{b1}) can be expressed in terms of explicit partial derivatives of $N_\alpha$, $N_\beta$, and the thermodynamic potential $\Omega$ with respect to $\mu_\alpha$, $\mu_\beta$, and $|\Delta_0|$ as
\begin{equation}\label{b2}
\frac{\tilde\kappa_{\alpha \beta} }{T} 
=
\left( \frac{\partial N_\alpha}{\partial \mu_{\beta}} \right) + 
\left( \frac{\partial N_\alpha}{\partial |\Delta_0|^2} \right)
\left( \frac{\partial N_{\beta}}{\partial |\Delta_0|^2} \right)
\left( \frac{\partial^2 \Omega}{\partial |\Delta_0|^4} \right)^{-1}
\end{equation}
From the expression above, we notice that $\tilde\kappa_{\alpha\beta} = \tilde\kappa_{\beta \alpha}$.

Once we have solutions of the order parameter equation together with the conservation of the particle number of each species for given values of $1/(k_{F+} a_s)$ and $P$, we can evaluate the pseudo-compressibility matrix elements $\tilde \kappa_{\alpha\beta}$ as a function of $1/(k_{F+} a_s)$ and $P$.
For instance, in the case of $s$-wave superfluid, the explicit expressions for Eq.~(\ref{b2}) are given by
\begin{equation}
\left( \frac{\partial N_\alpha }{\partial \mu_{\beta} } \right)
= \sum_{\bf k} 
\left[ 
\frac{|\Delta_0|^2}{4E_{}^3}
\,\mathcal X^-
- A_{\alpha\beta} (T) 
\right],
\end{equation}
\begin{equation}
\left( \frac{\partial N_\alpha }{\partial |\Delta_0|^2 } \right)
= \sum_{\bf k}  
\left[ 
\frac{\xi_{+}}{4E_{}^3} 
\,\mathcal X^-
+ 
\frac{B_{\alpha} (T)}{2E}
\right],
\end{equation}
\begin{equation}
\left( \frac{\partial^2 \Omega}{\partial |\Delta_0|^4} \right)
=
\sum_{\bf k}  
\left[
\frac{ 1  }{4E_{}^3} 
\,\mathcal X^-
+ 
\frac{C (T)}{4E^2}
\right].
\end{equation}
Here, we used the same notations 
$
\mathcal X^- =
n_F(\mathcal E_1) - n_F(\mathcal E_2)
$ as in Eq.~(\ref{eqn:fermi-fermi-order-parameter}).
And the second terms in the above equations are given by
\begin{equation}
A_{\alpha\beta} (T) = 
\sum_{\nu=1,2}
U_\alpha^\nu \,U_\beta^\nu
\left( \frac{n_F(\mathcal E_\nu)}{\partial \mathcal E_\nu} \right),
\end{equation}
\begin{equation}
B_{\alpha} (T) = 
\sum_{\nu=1,2}
U_{\alpha}^{\nu} \, 
\left( \frac{n_F(\mathcal E_\nu)}{\partial \mathcal E_\nu} \right),
\end{equation}
\begin{equation}
C (T) = 
\sum_{\nu=1,2}
\left( \frac{n_F(\mathcal E_\nu)}{\partial \mathcal E_\nu} \right),
\end{equation}
where 
$
U_\alpha^\nu =
\frac{1}{2}
\left(\gamma_\alpha(-1)^\nu+\xi_+/E\right)
$
with
$\gamma_{\uparrow} = 1$ and $\gamma_{\downarrow} = -1$.

At zero temperature, the Fermi distribution can be replaced by the step function as $n_F(\mathcal E_\nu) \to \theta(-\mathcal E_\nu)$, leading to
\begin{equation}
\mathcal X^- \vert_{T=0}
=
\theta(-\mathcal E_1) - \theta(-\mathcal E_2),
\end{equation}
and the corresponding the partial derivatives are by the Dirac delta function as $(\partial n_F(\mathcal E_\nu) / \partial \mathcal E_\nu ) \to - \delta(\mathcal E_\nu)$.
For convenience, let us introduce a dimensionless variable $z = (k/k_{F,+})^2$, $\tilde \mu_\pm = \mu_\pm / \epsilon_{F,+}$, and $\tilde \Delta_0 = \Delta_0 /\epsilon_{F,+}$, leading to $ \mathcal E_\nu (z)  = \mathcal E_\nu  / \epsilon_{F,+}$, where
\begin{equation}
\mathcal E_\nu (z) 
= 
\tilde m ( z - \tilde\mu_- )
+(-1)^\nu \,
\sqrt{ ( z- \tilde\mu_+)^2 
+ 
|\tilde\Delta_0|^2 }.
\end{equation}
Then, $\delta (\mathcal E_\nu) = \delta(\mathcal E_\nu (z)) / \epsilon_{F,+}$.
As a next step in the calculate, we use the relation
\begin{equation}
\delta(\mathcal E_\nu (z))
= \sum_{l} \frac{ \delta(z-z_{l})}{ \left| \mathcal E_\nu^\prime(z) \right|_{z_{l}}},
\end{equation}
where
$
\mathcal E_\nu^\prime(z)
= 
\tilde m +(-1)^\nu (z - \tilde\mu_+)/\sqrt{ (z - \tilde\mu_+)^2 + |\tilde\Delta_0|^2 }$,
and $z_{l}$ are the solutions of $\mathcal E_\nu(z) = 0$ given by
\begin{equation}
z_{l} 
= 
\frac{ \tilde\mu_+ -\tilde m \,\tilde\mu_- +(-1)^l \sqrt{D} }{1-\tilde m^2},
\end{equation}
where 
$
D = (\tilde\mu_+\,\tilde m  - \tilde\mu_- )^2 - (1-\tilde m^2) |\tilde\Delta_0|^2$, and $\tilde m = (1 - m_r)/(1+m_r) $ with $m_r =m_\uparrow / m_\downarrow$ the mass ratio of the spin-up to spin-down particles.

Now, we can evaluate $A_{\alpha\beta}(0)$, $B_\alpha(0)$, and $C(0)$ by considering the signs of $D$ and $z_l$.
For instance, if $D < 0$, $A_{\alpha\beta}(0) = B_\alpha(0) = C(0) = 0$.
But, if $D\ge 0$ and $z_l \ge 0$, then we have
\begin{equation}
\sum_{\bf k} A_{\alpha \beta} (0)
= 
\frac{3N_+}{4\epsilon_{F,+}} 
\sum_{\nu,l}
\sqrt{z_{l}} 
\Big[\,
a_{\alpha \beta} (z_{l}) \,
W_\nu(z_l) 
\,\Big],
\end{equation}
\begin{equation}
\sum_{\bf k} \frac{ B_\alpha (0) }{ 2 E }
= 
\frac{3N_+}{4\epsilon_{F,+}} 
\sum_{\nu,l}
\sqrt{z_{l}}
\Big[\,
\frac{b_{\alpha}(z_{l})}{2E(z_{i})} \,
W_\nu(z_l)
\,\Big],
\end{equation}
\begin{equation}
\sum_{\bf k} \frac{ C(0) }{ 4 E^2 }
= 
-\frac{3N_+}{4\epsilon_{F,+}} 
\sum_{\nu,l}
\sqrt{z_l}
\Big[\, 
\frac{1}{4E^2(z_{l})}\,
W_\nu(z_l)
\,\Big].
\end{equation}
where
$
a_{\alpha \beta} (z_{l})
= 
U_\alpha^{\nu}(z_{l}) \, 
U_\beta^{\nu} (z_{l})
$,
$
b_{\alpha}(z_{l})
=
U_\alpha^\nu (z_{l})
$, and
$
W_\nu(z_{l}) 
=
1/\left| \mathcal E_\nu^\prime (z_l) \right|
$.

\section{Analytical Result of $(\partial N / \partial \mu)_{T}$ for the Equal Mass Case at $P=0$ and $T=0$ }
\label{app-c}
In this appendix, we present analytical results of isothermal susceptibility  $(\partial N / \partial \mu)_T$ for equal mass and population case at $T=0$ in both BCS and BEC regimes.
The total number of particles $N = N_\uparrow + N_\downarrow$ at $T=0$ is 
\begin{equation}
N = \sum_{\bf k} \left[ 1 - \frac{\xi_{\bf k}}{E_{\bf k}} \right],
\end{equation}
where $\xi_{\bf k} = k^2 / 2m - \mu$ and $E_{\bf k} = \sqrt{ \xi_{\bf k}^2 + |\Delta_{\bf k}|^2}$. 
For $s$-wave superfluid, $\Delta_{\bf k} = \Delta_0$.

\subsection{BCS limit}
In the BCS regime, it is known that $
\tilde \Delta_0 \equiv \Delta_0 / \epsilon_F \simeq (8/e^2) \, e^{-(\pi/2) |\lambda|}
$ and $
\mu / \epsilon_F \simeq 1
$ with $
\lambda = 1/(k_F a) \to -\infty
$~\cite{engelbrecht-97}, leading to $
(\Delta / \mu ) \ll 1,
$ and
\begin{equation}
\label{delta-deritave}
\frac{\partial |\Delta_0|^2}{\partial \mu}  = \frac{2|\Delta_0|^2}{\mu}.
\end{equation}
Now the susceptibility $\chi = ( \partial N / \partial \mu)$ can be evaluated using Eq.~(\ref{delta-deritave}), 
\begin{eqnarray}\nonumber
\frac{\partial N}{\partial \mu} 
&=& 
\left(\frac{\partial N}{\partial \mu}\right)
+
\left(\frac{\partial N}{\partial |\Delta_0|^2} \right)
\left(\frac{\partial |\Delta_0|^2}{\partial \mu}\right) 
\\
&=&
\sum_{\bf k} \left[ \frac{1 + \xi_{\bf k} / \mu }{E_{\bf k}^3} \right] 
|\Delta_0|^2.
\end{eqnarray}
Here, we introduce a dimensionless variable $x=(\epsilon_F / \mu) (k/k_F)^2$, leading in thermodynamic limit to the summation over momentum
\begin{equation}
\sum_{\bf k} = V \int \frac{d^3 k}{(2\pi)^3} =  C \int_0^\infty dx\, \sqrt{x} ,
\end{equation}
where $
C = (3N/4)\, (\mu/\epsilon_F)^{3/2} \simeq (3N/4).
$
Then, we have
\begin{equation}
\frac{\partial N}{\partial \mu} 
=
\frac{3N}{4\epsilon_F} \,
|\tilde \Delta_0|^2
\int_0^\infty dx \, 
\left[
\frac{x}
{ (x-1)^2 + |\tilde \Delta_0|^2 } \right]^{3/2} 
\end{equation}
By changing the variable $x$ to $x+1$ and separating the integration range in two parts
\begin{equation}\label{sus-bcs}
I_1 + I_2 = 
\left( 
\int_{-1}^1 dx 
+
\int_{1}^\infty dx 
\right) 
\,
\left[\,
\frac{ x+1 }
{ x^2 + |\tilde \Delta_0|^2 } 
\,\right]^{3/2},
\end{equation}
we can perform this integration.
The integrand in the first region becomes $
1/ (\, x^2 + |\tilde \Delta_0|^2 \,)^{3/2} ,
$ in the $|\tilde \Delta_0| \ll 1$ limit.
Noticing that $\int_{-1}^{1} \,dx \, (x^2+d^2)^{-3/2} = (2/d^2) / \sqrt{1+d^2}$, the first integral is 
\begin{equation}\label{integration-1}
I_1 
\simeq 
\frac{2}{ |\tilde \Delta_0|^2 }
\left[ 1 - \frac{|\tilde \Delta_0|^2}{2} 
\right].
\end{equation}
The integrand in the second region can be approximated by $
(x+1)^{3/2} / x^3 ,
$ since $\vert \tilde \Delta_0 \vert \ll 1$, leading to 
\begin{equation}\label{integration-2}
I_2 = 
\int_{1}^\infty dx \,
\frac{ (x+1)^{3/2} }
{ x^3 } =
\frac{1}{4} \Big[ 7\sqrt{2} + 3 \sinh^{-1}(1) \Big]
\end{equation}
Substituting Eq.~(\ref{integration-1}) and (\ref{integration-2}) into Eq.~(\ref{sus-bcs}), we have $(\partial N / \partial \mu)$ in the BCS regime 
\begin{equation}
\frac{\partial N}{\partial \mu}
\simeq
\frac{3N}{2\epsilon_F}
\Big[ 1 + \gamma \,e^{-\pi / (k_F |a|) } \Big],
\end{equation}
where
$\gamma = [7\sqrt{2} - 4 + 3 \sinh^{-1}(1) ] 8/e^4 \simeq 1.2519$.

\subsection{BEC limit}
Next, we consider the BEC regime, where $k_F a_s \to  0^{+}$.
It is known~\cite{engelbrecht-97} that the order parameter $
|\Delta_0| = (16/3\pi)^{1/2} \epsilon_F\, / \sqrt{k_F a_s}
$ and chemical potential $
\mu = -E_b/2 + (2 / 3\pi) \epsilon_F (k_F a_s),
$ with $E_b = 1/(ma_s^2)$ a binding energy, leading to 
\begin{equation}
|\Delta_0|^2 
= 
\frac{4}{3\pi} 
\frac{k_F^3}{m^2 a_s}
\end{equation} 
and
\begin{equation}
\mu 
= 
-
\frac{E_b}{2} 
+ 
\frac{2a_s}{3\pi} 
\frac{k_F^3}{2m}.
\end{equation}
For a given scattering length $a_s$, we can perform the partial derivative of $|\Delta_0|^2$ with respect to $\mu$, by parametrizing the chemical  via $k_F^3 \sim N/V$, leading to
\begin{equation}
\frac{\partial |\Delta_0|^2}{\partial \mu} 
=
\left( \frac{\partial |\Delta_0|^2}{\partial k_F^3} \right) /
\left(  \frac{\partial \mu}{\partial k_F^3} \right)
=
\frac{4}{ma_s^2}.
\end{equation}
Taking $\lambda = 1/(k_F a_s ) \gg 1$, we can rewrite $|\Delta_0|$ and the corresponding derivative as
\begin{equation}
|\Delta_0|^2 
= 
\frac{16}{3\pi} \epsilon_F^2 \lambda,
\quad
\frac{\partial |\Delta_0|^2}{\partial \mu}  
= 8 \epsilon_F \lambda^2 .
\end{equation}
Introducing the dimensionless variable $x = (k/k_F)^2 / \lambda^2  $, in the thermodynamic limit, the momentum sum reads 
\begin{equation}
\sum_{\bf k} = \frac{3N}{4} \lambda^3 \int_0^\infty dx \,\sqrt{x}.
\end{equation}
Notice that the chemical potential in this large $\lambda$ limit is of the order of $\lambda^2$, since $
E_b / 2 = \epsilon_F \lambda^2,
$ and  
\begin{equation}\nonumber
\xi_{\bf k} \simeq \epsilon_F \lambda^2 ( x + 1),
\end{equation}
\begin{equation}\nonumber
E_{\bf k} = \epsilon_F \lambda^2 \sqrt{ 
(x+1)^2 + (16/3\pi)/\lambda^3 } 
\simeq \xi_{\bf k} + \mathcal{O}(\lambda^{-1}).
\end{equation}

Next, we consider the order of magnitude of the explicit and implicit derivatives. 
First, we notice that the explicit derivative of $|\Delta_0|$ with respect to $\mu$ is of the order of $\lambda^{-2}$ 
\begin{equation}
\left(\frac{\partial N}{\partial \mu}\right)_{\rm{ex}}
=
\sum_{\bf k} \frac{|\Delta_0|^2}{E^3}
\sim 
\mathcal{O}(\lambda^{-2}).
\end{equation}
Second, we establish that the implicit derivative is of the order of $\lambda$ 
\begin{equation}
\left(\frac{\partial N}{\partial \mu} \right)_{\rm{im}}
=
\sum_{\bf k} \left( \frac{\xi_{\bf k}}{2E_{\bf k}^3}\right)
\left( \frac{\partial \Delta^2}{\partial \mu} \right)
\sim
\mathcal{O} (\lambda).
\end{equation}
Thus, in the BEC regime $\lambda \gg 1$, $\partial N / \partial \mu$ can be approximated by the leading order term of the implicit derivative:
\begin{equation}
\frac{\partial N}{\partial \mu}
\simeq
\frac{3N}{4}
\frac{4}{\epsilon_F} \lambda \int_0^\infty dx\, \sqrt{x}
\,(x+1)^{-2}.
\end{equation}
Noticing that $
\int_0^\infty dx \,\sqrt{x}\,(x+1)^{-2} 
= \pi/2
$, we have
\begin{equation}
\frac{\partial N}{\partial \mu}
\simeq
\frac{3N}{2 \epsilon_F}
\frac{\pi}{k_F a}.
\end{equation}
Since the total number of particles is $N = N_\uparrow + N_\downarrow = k_F^3/(3\pi^2)$ and the Fermi energy is $\epsilon_F = k_F^2/2m$, we can express the derivative
\begin{equation}\label{fermi-sus}
\frac{1}{V}\,\left(\frac{\partial N}{\partial \mu}\right)_T
=
\frac{1}{\pi}\,\left(\frac{m}{a_s}\right),
\end{equation}
in terms of the mass and the scattering length of the fermions.
Since in the BEC limit all fermions are paired into tightly bound molecules a direct comparison to the results of weakly interaction Bose gas is possible.
For a weakly interacting bosons, it is known~\cite{huang-book} that
\begin{equation}\label{boson-sus}
\frac{1}{V} \, \left(\frac{\partial N_B}{\partial \mu_B} \right)_T
=
\frac{1}{4\pi}\,\left( \frac{m_B}{a_B}\right)
\end{equation}
in their superfluid state at $T =0$.
Given that the number of bosons is $N_B = N/2$, and that $(\partial \mu_B)/(\partial \mu) \simeq 2$, a substitution of the boson mass $m_B = 2m$ and its effective scattering length $a_B = 2a_s$ into Eq.~(\ref{fermi-sus}), leads to $(m/a) = (m_B/a_B)$. Thus, the density susceptibility for a Fermi superfluid in the BEC limit recovers the the expected results for weakly interacting Bose-Einstein condensates and lead to the relation:
\begin{equation}
\left(\frac{\partial N_B }{ \partial \mu_B } \right)_T
=
\frac{1}{4} \,\left(\frac{\partial N }{ \partial \mu } \right)_T.
\end{equation}


\begin{thebibliography}{99}

\bibitem{ketterle-10}
C. Sanner, E. J. Su, A. Keshet, R. Gommers, Y. Shin, W. Huang, 
and W. Ketterle,
Phys. Rev. Lett. {\bf 105}, 040402 (2010).


\bibitem{esslinger-10}
T. M{\"u}ller, B. Zimmermann, J. Meineke, J.-P. Brantut,
T. Esslinger, and H. Moritz,
Phys. Rev. Lett. {\bf 105}, 040401 (2010).


\bibitem{chin-10}
C. L. Hung, X. Zhang, N. Gemelke, and C. Chin,
Nature {\bf 470}, 236 (2011).


\bibitem{iskin-05}
M. Iskin and C. A. R. S\'a de Melo,
Phys. Rev. B {\bf 72}, 224513 (2005).


\bibitem{iskin-06a}
M. Iskin and C. A. R. S\'a de Melo,
Phys. Rev. A {\bf 74}, 013608 (2006).


\bibitem{raizen-05}
C.-S. Chuu, F. Schreck, T. P. Meyrath, J. L. Hanssen, G. N. Price, and M. G. Raizen,
Phys. Rev. Lett. {\bf 95}, 260403 (2005).


\bibitem{bloch-05}
S. F\"olling, F. Gerbier, A. Widera, O. Mandel, T. Gericke, and I. Bloch,
Nature, {\bf 434}, 481 (2005).


\bibitem{bouchoule-06}
J. Esteve, J.-B. Trebbia, T. Schumm, A. Aspect, C. I. Westbrook, and I. Bouchoule,
Phys. Rev. Lett. {\bf 96}, 130403 (2006).


\bibitem{steinhauer-10}
A. Itah, H. Veksler, O. Lahav, A. Blumkin, C. Moreno, C. Gordon, and J. Steinhauer,
Phys. Rev. Lett. {\bf 104}, 113001 (2010).


\bibitem{stringari-11}
M. Klawunn, A. Recati, L. P. Pitaevskii, and S. Stringari,
arXiv:1102.3805 (2011).


\bibitem{ketterle-06}
M. W. Zwierlein, A. Schirotzek, C. H. Schunck, 
and W. Ketterle,
Science {\bf 311}, 492 (2006).


\bibitem{hulet-06}
G. B. Partridge, W. Lui, R. I. Kamar, Y. Liao, and
R. G. Hulet,
Science {\bf 311}, 503 (2006).


\bibitem{grimm-10}
F. M. Spiegelhalder, A. Trenkwalder, D. Naik,
G. Kerner, E. Wille, G. Hendl, F. Schreck, 
and R. Grimm,
Phys. Rev. A {\bf 81}, 043637 (2010).


\bibitem{grimm-11}
A. Trenkwalder, C. Kohstall, M. Zaccanti, D. Naik, A. I. Sidorov, F. Schreck, and R. Grimm,
Phys. Rev. Lett. {\bf 106}, 115304 (2011).


\bibitem{ketterle-11}
C. Sanner, E. J. Su, A. Keshet, W. Huang, J. Gillen, R. Gommers, and W. Ketterle,
Phys. Rev. Lett. {\bf 106}, 010402 (2011)


\bibitem{seo-11}
Kangjun Seo, and C. A. R. S\'a de Melo,
arXiv:1101.3610 (2011).


\bibitem{kubo-57}
R. Kubo, 
Rep. Prog. Phys. {\bf 29}, 255 (1966);
M. Toda, R. Kubo,
Statistical Physics, 2nd ed. (1991)


\bibitem{botelho-05}
S. S. Botelho and C. A. R. S\'a de Melo,
J. Low Temp Phys. {\bf 140}, 409 (2005).


\bibitem{yip-06}
C. H. Pao, S.-T. Wu, and S. K. Yip, 
Phys. Rev. B {\bf 73}, 132506 (2006).


\bibitem{mueller-06}
T. N. De Silva and E. J. Mueller,
Phys. Rev. A {\bf 73} 051602(R)


\bibitem{iskin-06b}
M. Iskin and C. A. R. S\'a de Melo,
Phys. Rev. Lett. {\bf 97}, 1000404 (2006).


\bibitem{stoof-06}
M. Haque and H. T. C. Stoof,
Phys. Rev. A {\bf 74} 011602(R) (2006).


\bibitem{pieri-06}
P. Pieri and G. C. Strinati,
Phys. Rev. Lett. {\bf 96}, 150404 (2006).


\bibitem{tempere-07}
J. Tempere, M. Wouters, and J. T. Devresse,
Phys. Rev. B {\bf 75}, 184526 (2007).


\bibitem{iskin-07}
M. Iskin and C. A. R. S\'a de Melo,
Phys. Rev. A {\bf 76}, 013601 (2007).


\bibitem{huang-book}
K. Huang and C. N. Yang,
Phys. Rev. {\bf 105}, 776 (1957);
K. Huang,
Statistical Mechanics (1987).


\bibitem{engelbrecht-97}
J. R. Engelbrecht, M. Randeria, and C. A. R. S\'a de Melo,
Phys. Rev. B {\bf 55}, 15153 (1997).


\end{thebibliography}
\end{document}